\documentclass[10pt,journal,compsoc]{IEEEtran}

\ifCLASSOPTIONcompsoc
  \usepackage[nocompress]{cite}
\else
  \usepackage{cite}
\fi

\ifCLASSINFOpdf
   \usepackage[pdftex]{graphicx}
\else
\fi

\usepackage{underscore}
\usepackage[document]{ragged2e}
\usepackage[ruled,vlined]{algorithm2e}
\usepackage{dblfloatfix}
\usepackage{caption}
\usepackage{hyperref}
\usepackage{amsmath}
\usepackage[all]{hypcap}
\usepackage[T1]{fontenc}
\DeclareUnicodeCharacter{0301}{\'{e}}

\begin{document}
\title{Toward Blockchain-Enabled Supply Chain Anti-Counterfeiting and Traceability}
\author{Neo~C.K.~Yiu,~\IEEEmembership{Member,~IEEE\\ Department~of~Computer~Science,~University~of~Oxford\\neo-chungkit.yiu@kellogg.ox.ac.uk}
\IEEEcompsocitemizethanks{\IEEEcompsocthanksitem Neo C.K. Yiu was with Department of Computer Science, University of Oxford, Oxford, United Kingdom}
\thanks{Manuscript first submitted for preprint on Jan 29, 2021.}}

\IEEEtitleabstractindextext{
\justify
\begin{abstract}
Innovative solutions addressing product anti-counterfeiting and record provenance have been deployed across today's internationally spanning supply chain networks. These product anti-counterfeiting solutions are developed and implemented with centralized system architecture relying on centralized authorities or any form of intermediaries. Vulnerabilities of centralized product anti-counterfeiting solutions could possibly lead to system failure or susceptibility of malicious modifications performed on product records or various potential attacks to the system components by dishonest participant nodes traversing along the supply chain. Blockchain technology has progressed from merely with an use case of immutable ledger for cryptocurrency transactions to a programmable interactive environment of developing decentralized and reliable applications addressing different use cases globally. In this research, so as to facilitate trustworthy data provenance retrieval, verification and management, as well as strengthening capability of product anti-counterfeiting in supply chain industry, key areas of decentralization and feasible mechanisms of developing decentralized and distributed versions of these product anti-counterfeiting and traceability ecosystems utilizing blockchain technology, are identified via a series of security and threat analyses performed mainly against the NFC-Enabled Anti-Counterfeiting System (NAS) which is one of these solutions currently implemented in supply chain industry with centralized architecture. A set of fundamental system requirements are set out for developing a blockchain-enabled autonomous and decentralized ecosystem for supply chain anti-counterfeiting and traceability, as a secure and immutable scientific data provenance tracking and management platform in which provenance records, providing compelling properties on data integrity of luxurious goods, are recorded and verified automatically, for supply chain industry.
\end{abstract}

\begin{IEEEkeywords}
Blockchain, Anti-counterfeiting, Decentralization, Product Authenticity, End-to-End Traceability, Supply Chain Integrity, Supply Chain Provenance, NFC-enabled Anti-counterfeiting System, Near-field Communication, Internet-of-Things.
\end{IEEEkeywords}}

\maketitle
\IEEEdisplaynontitleabstractindextext
\justifying
\IEEEpeerreviewmaketitle

\ifCLASSOPTIONcompsoc
\IEEEraisesectionheading{\section{Introduction}\label{sec:introduction}}
\else
\section{Introduction}
\label{sec:introduction}
\fi

\IEEEPARstart{I}{n} this chapter, the motivation of this research, which is mainly based on the current challenges in supply chain anti-counterfeiting and traceability, and a list of research objectives with the preferred research methodology are also pinpointed by going through predefined individual research questions so as to answer the growing challenges of improving the worsening situations in counterfeit product trading. 

\subsection{The Current Challenges in Supply Chain Anti-Counterfeiting}
The problem of counterfeit product trading, including luxurious goods or pharmaceutical products, has been one of the major challenges the supply chain industry has been facing, in an innovation-driven global economy. The situation has exacerbated with an exponential growth of counterfeits and pirated goods worldwide, for which it has also plagued companies with multinational supply chain networks for decades and on. 

The analytical study – \cite{1}, published by OECD (Organization for Economic Cooperation and Development) and EUIPO (European Union Intellectual Property Office) in 2019 regarding the global trading activities of counterfeit and pirated products, has suggested the volume of international trade of counterfeit and pirated products increased from \$250 billion in 2007, to up to \$461 billion in 2013 representing approximately 2.5\% of world imports, of which the imports of counterfeit and pirated products into EU amounted to nearly \$116 billion representing to 5\% of EU imports approximately. The results have been alarming and further exacerbating according to the latest statistics published in 2016, suggested that the volume has already further amounted to as much as \underline{\emph{\$509 billion}} representing \underline{\emph{3.3\%}} of world trade and 6.8\% of imports from non-EU countries.

The battle against counterfeit product trading remains a significant challenge, and simultaneously this is of significant and growing concern to the globalized economy not to mention all the affected industries such as the markets included in \cite{2} and innumerable branded-product companies. Trading activities in counterfeit not only infringe on trademarks and copyrights, negatively impact on sales and profits of industries but also generate profits for organized crimes at expense of the affected companies and governments, as well as posing broader adverse effects to the economy, public health, safety and security of the wider community. Counterfeit product trading activities are operated swiftly in the globalized economy, misusing free trade zones according to \cite{3}, taking advantage of many legitimate trade facilitation mechanisms and thriving in economies with weak governance standard and limited innovative options to combat product counterfeits.

In response to the growing concern, innovative, fully-functional, integrable and affordable product anti-counterfeiting solutions with traceability functionalities, utilizing the cutting-edge technologies, have been widely and urgently demanded so as to ensure the provenance and traceability of genuine products throughout the supply chain counterfeiting, and these suggested solutions should be widely adopted regardless of industries, size of the companies and its supply chain systems.

\subsection{Research Objectives}
Open research questions and difficulties inherent in addressing them should first be detailed before actually diving into foundation and latest developments of Blockchain technology. It will then be followed by having a brief overview on one of the existing product anti-counterfeiting systems - \emph{NFC-Enabled Anti-Counterfeiting System (NAS)}, and performing security analyses on NAS to gather insightful findings on opportunities of how these product anti-counterfeiting solutions could be improved with decentralization enabled by Blockchain technology. 

Given the relentless and exacerbating counterfeit trading activities though there has already been a variety of innovative product anti-counterfeiting solutions introduced in supply chain industry, the main research question should therefore be:

\emph{"Why would existing anti-counterfeiting and traceability systems benefit from decentralization enabled by blockchain technology to better combat the rampant counterfeiting attacks?"}

The main question could further be addressed via answering the following sub-questions throughout the research:
\begin{enumerate}
  \item Why blockchain could be utilized to decentralize the supply chain anti-counterfeiting and traceability system?
  \item What are the security limitations and concerns of existing supply chain anti-counterfeiting and traceability systems, such as NAS?
  \item What are the advantages introduced by decentralized supply chain anti-counterfeiting and traceability systems?
\end{enumerate}

Given the main research question and a set of sub-questions derived, it is common to follow an organised way of exploring them step-wise in this research, for which an insightful background on blockchain technology and NAS are given, followed by a series of security analyses which could ultimately produce evidence-based answers to the research questions, with discussion on what and how a decentralized supply chain anti-counterfeiting and traceability system, using blockchain technology and other related technical concepts, could improve the exacerbating situation of product counterfeits.

\section{Background}
Given the growing concern on trading activities of counterfeit products, there have been anti-counterfeiting solutions developed and implemented in supply chain systems of different industries. The contributions of this research are also proposed and elaborated in this chapter.

\subsection{Related Work of Anti-Counterfeiting and Traceability Systems}
The starting point, such as \cite{4} is always digitalizing the multi-node supply chain system under which the supply chain network, will at least be operating with a POS (Point-of-Sale) system integrated with states of data storage system updated at different nodes along the supply chain. Wireless communication technologies, such as RFID (Radio-frequency Identification) or NFC (Near-field communication which is a subset of RFID), powering the Internet-of-Things (IoT) are mostly the innovative solutions with centralized architectures currently based on, via packaging the tags on the packets of goods or the product itself for identification, anti-counterfeiting and traceability purposes. 

RFID-based solutions in \cite{7,8,9,10} and NFC-based solutions in \cite{22,23,24,25} both require dedicated applications or authentication servers to integrate with these tags involved so as to perform writing, reading and validating features on the data stored in these tags with the product identifier (PID) and its metadata also stored in database system of applications for the purpose of validation at later stage to respond to any potential counterfeiting attack at different supply chain nodes. \cite{11, 12} utilized simple barcode technology such as QR code to retrieve the product identifiers by scanning the products and querying corresponding records throughout the dedicated backend database systems to retrieve metadata of specific PIDs so as to prove the authenticity with the PIDs validated and detailed product data retrieved.

For instance, with the wider adoption of these wireless communication and barcode technologies, the concept of RFID-enabled track-and-trace anti-counterfeiting was also proposed in \cite{13,14,15,16} to combat counterfeiting activities. There are RFID-enabled anti-counterfeiting approaches further analysed and compared, with potential implementation issues also discussed in \cite{10,17}. \cite{18,19,20} further extended the EPCglobal traceable system to support anti-counterfeiting. \cite{21} suggested the use of Wireless Sensor Network (WSN) and geospatial technologies, such as Geographic Information System (GIS), Global Positioning System (GPS), Remote Sensing (RS) for mobile assets-tracking operations, are amongst technological innovations that have been applied in product traceability systems.

Nevertheless, many of the existing implementations for product anti-counterfeiting in supply chain industry, utilizing wireless communication technologies, wireless sensor networks or geospatial technologies, are built with centralized architecture relying on a trusted server, database and applications, which are solely controlled and managed by the manufacturers or suppliers of the products, for coordinating and managing product authentication with participation contributed from different nodes along the supply chain of different industries.

\subsection{Research Contributions}
Contrary to existing conceptual blockchain implementations applied in different industries with less explanation on why decentralization is needed for these use cases, this research is taking a rather different approach to have a thorough process involving a series of security analyses on an existing product anti-counterfeiting and traceability system - NAS, to explain why these product anti-counterfeiting and traceability solutions could be benefit from decentralization, on which a decentralized version of these supply chain software solutions could be developed and implemented. 

The fundamental ideas and use cases of blockchain technology, categorized into different phases, are explained in the research. The reasons why implementations of Blockchain 2.0 should be adopted to decentralize the existing product anti-counterfeiting and traceability systems are also elaborated, before stepping into the overview on one of the existing product anti-counterfeiting and traceability system - NAS, for which this research will focus on so as to explain the reasons why these existing supply chain software solutions including NAS would be benefit from decentralization enabled by blockchain technology.

Based on the findings and opportunities identified from these analyses, a set of fundamental system requirements of developing a decentralized product anti-counterfeiting and traceability system is also defined. The decentralized solutions, such as the Decentralized NFC-Enabled Anti-Counterfeiting System (dNAS), are aimed at delivering a more secured and quality approach to verify authenticity and provenance of luxurious products such as bottled wine, with the use of peer-to-peer blockchain networks and distributed storage technologies to eliminate absurd amount of costs for on-chain storage and provide a much higher level of privacy, reliability and quality of service compared with existing anti-counterfeiting and traceability solutions with centralized architecture. The decentralized version of the existing solutions, such as dNAS, could also define a framework and practice for different nodes along the supply chain to integrate the low-cost, real-time and immutable blockchain technology into their daily supply chain workflows. 

\section{Decentralizing with The Blockchain Technology}
In this chapter, following the advantages introduced by the blockchain technology, the modern core blockchain concepts including those of blockchain 1.0 and blockchain 2.0 are explained. A variety of current blockchain implementations applied to different types of supply chain industries are also mentioned with reasons on why developing decentralized solutions based on existing systems of centralized architecture is a pragmatic way going forward to improve the challenging product counterfeiting in the wider supply chain industry.

\subsection{Blockchain At the Core}
The decentralized architecture could bring more advantages to the existing centralized product anti-counterfeiting system, and an example could be decentralizing NAS utilizing blockchain technology. Blockchain is a distributed ledger technology recording and sharing all the transactions that occur within a dedicated peer-to-peer network. It is essentially a decentralized timestamp service with a virtual machine to execute signed scripts that operates on signed data. It utilizes distributed ledger to store scripts and data with mutual consensus reached amongst participating nodes running on the same blockchain network.

\begin{figure}[h]
    \centering
    \captionsetup{justification=centering}
    \includegraphics[width=0.5\textwidth]{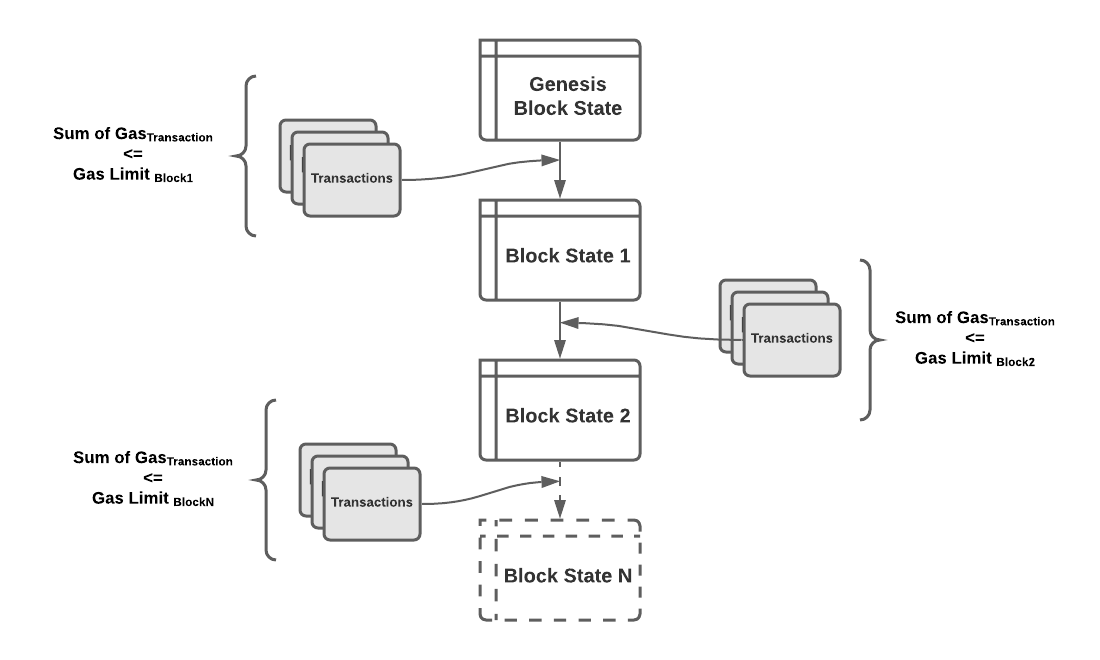}
    \caption{\textit{Block State with Transactions Flow}}
    \label{fig:blockstate}
\end{figure}

The blockchain network is consisted of multiple nodes that maintain a set of shared states and perform transactions updating the states which could be divided into ledger state, block state and transaction state as depicted in \textit{Fig.~\ref{fig:blockstate}}. Transactions need to be going through the mining process, in which it must be validated by the majority or agreed fraction amongst the participating network nodes, depending on which consensus protocol is adopted, before being ordered and packaged into a timestamped block which is also known as \emph{block signing}.

The blockchain network can be generally categorized either as permission-less (public network) or permissioned network. The former is an open distributed ledger network such as \cite{28}, where any node can join the network and where any two peers can conduct transactions without any authentication performed by any central authority, while the latter is a controlled distributed ledger like \cite{30} where the decision making and validation process are kept to one organization or few organizations forming a consortium with or without the staking concept. In permissioned networks, the consortium administrator or certificate authority determines who can join the network as a validator node or listener node, if there is no logic of on-chain governance available. All nodes are authenticated in advance, and their identity is known to other nodes running on the same network and in the same consortium, at least to the administrator.

\begin{figure}[h]
    \centering
    \captionsetup{justification=centering}
    \includegraphics[width=0.5\textwidth]{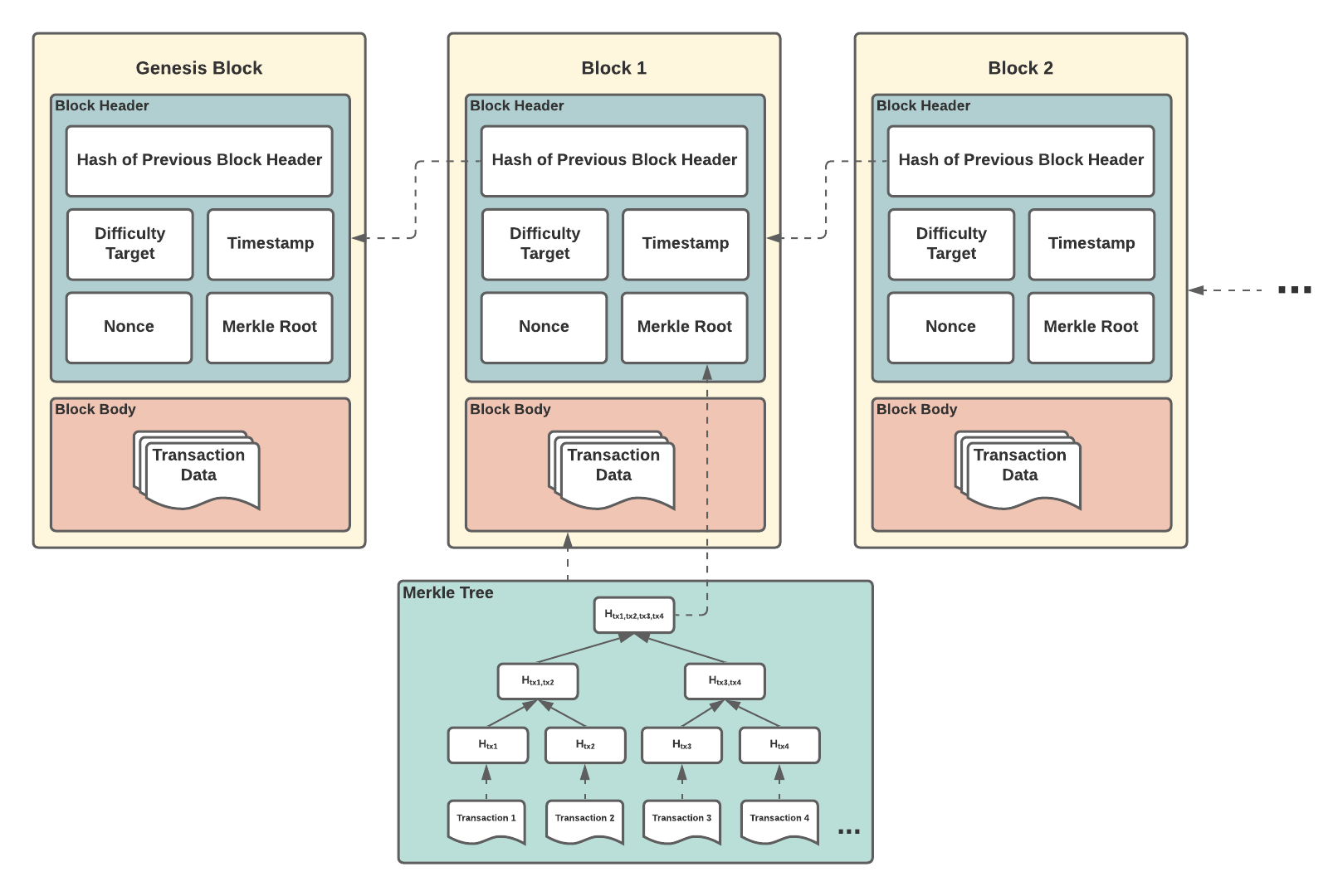}
    \caption{\textit{General Blockchain Data Structure}}
    \label{fig:generalblockchains}
\end{figure}

The general blockchain data structure is demonstrated in \textit{Fig.~\ref{fig:generalblockchains}}. The first block is always referred as \emph{genesis block}, and a block is consisted of a header and a body. The block body contains the list of transactions. The number of transactions that could be fit into a block is dependent on the block size (block gas limit) and the transaction size (gas spent per transaction). The block header contains a wide variety of fields, ranging from timestamp, Merkle root hash representing the hash value of every transaction in the block, the hash pointer of the previous block header for which different blocks are "\emph{chained}" to each other by putting this field of hash for every next block, the nonce which is the 32-bit field incremented until the equation is solved, to the difficulty which is needed for the \emph{Proof-of-Work (PoW)} protocol which is heavily linked with computation process known as \emph{mining}, for which miners are the nodes to calculate the block header hash termed as "\emph{solving the puzzle}".

\subsection{Characteristics of Blockchain}
Based on the Proof-of-Work consensus protocol, the block is said to be mined if a miner finds its nonce such that the hash of block header is less than the difficulty target, based on \cite{28}. Modern blockchain is also characterized into FOUR main aspects, apart from being utilized merely as the distributed ledger, namely the smart contract, immutability, cryptography and consensus:

\underline{\emph{Smart contracts}} are self-executing scripts with predefined methods deployed and stored on blockchain. Smart contract methods are invoked to execute accordingly on every node of the blockchain network when performing a transaction. Every node of the blockchain network must agree on the inputs, outputs and states transited by the smart contract. Common contractual conditions, as discussed in \cite{36, 37, 38}, such as payment terms and conditions could be satisfied, thus minimizing the need for trusted intermediaries but the decentralized “code” instead.

\underline{\emph{Immutable in nature}}, meaning that there is one, universally accepted version of the chain history which cannot be easily altered which is termed as blockchain state, or transaction states adding up to the block state and in turn the global ledger state.

\underline{\emph{Cryptographic techniques}} are used in the blockchain to ensure authenticity, integrity, immutability and non-repudiation of the blockchain networks owing to the fact that even an authenticated node can act maliciously in the network. The state root hash and the hash pointers are in combination to secure and track every historical change made to the global ledger state. The purpose of having Merkle root hash of every block is to detect data tampering and to validate transactions effectively and efficiently with the Merkle proof technique. Every transaction hash is validated by going through the Merkle tree path to see if an identical root hash can be obtained, as such any modification of the transaction in the block and its transaction hash will be instantly detected since the root hash obtained will be different. The block header root hash is to verify the integrity of the block and its transactions, and link to the previous block by also embedding its block root hash in the current block header so as to form a "\emph{chain}". The asymmetric cryptography is also adopted, and every pending transaction is needed to be signed before being broadcast across the blockchain network. Each account of the blockchain node has a set of key pair of which the private key is used to encrypt the hash value derived from the transactions and the public key is used by the peer node to determine the authenticity and integrity of transactions.

\underline{\emph{A consensus}} is an agreement of a decentralized network to authenticate and validate a transaction, after which every node shares and stores the same data and state to its persistent storage so as to prevent the malicious actors from manipulating the data like the situation of those software solutions with traditional and centralized client-server architecture. There are a few parameters to consider a consensus mechanism, namely the authentication, data integrity, byzantine fault tolerance, decentralized governance, quorum structure, performance, non-repudiation. For instance, Proof-of-Work (PoW) is a computation-expensive mining protocol to work around the Sybil attack where a minority can control the whole network and manipulate the double-spend problem. Proof-of-Stake (PoS) is more focused on scalability and efficiency of the network handling block mining and transaction validation. Regarding permissioned blockchain, it is all down to the organization or consortium hosting majority of nodes running on the blockchain network to determine its consensus protocol and process.

\subsection{Starting from the Original Blockchain 1.0 – The Bitcoin Network}
Blockchain is often regarded as the underlying technology of Bitcoin \cite{28} – peer-to-peer version of electronic cash, namely the decentralized virtual currency which does not require any existing currency institutions to circulate and is of fixed currency circulation. Bitcoin network is indeed the first use case adopting blockchain technology. Bitcoin aimed at offering a purely peer-to-peer version of electronic cash which would allow online payments to be sent directly from one party to another without involving a financial institution. The main benefits of such decentralized virtual currency system are to prevent from the double spending, single point of control and potential failure due to the reliance of trusted third parties and intermediaries. Bitcoin network relies heavily on decentralized consensus and its cryptographic properties with use of digital signature instead, offering new transparency to finance industry, which have normally been of great security concerns on virtual currencies.

Blocks of the Bitcoin network are mined through a computationally-intensive process also known as the Proof-of-Work consensus protocol requiring significant computational resources to solve a cryptographic hash-based puzzle and the solution could be worked out by trial-and-error based on the targeted difficulty set per block. The consensus must be reached before a new block could be created with respective transactions packed in the new block. As there are many miner nodes available on the open Bitcoin network, every miner on the network is literally competed with each other to generate a valid Proof-of-Work consensus for the block, and it will take roughly 10 minutes on average with the current setting of Bitcoin network to have a miner created a block successfully and received the mining reward which has been halved on predefined milestone blocks (also known as the "\emph{halving}") of Bitcoin network. The Proof-of-Work adopted in Bitcoin network would prevent the Sybil attackers from promoting a dishonest blockchain supporting their malicious agendas, offering a way for honest nodes to overcome Byzantine failures as well as accepting the next block on the canonical chain, which is arguably the most difficult part of implementing a consensus protocol where many attack vectors would be focused on.

\begin{figure}[h]
    \centering
    \captionsetup{justification=centering}
    \includegraphics[width=0.5\textwidth]{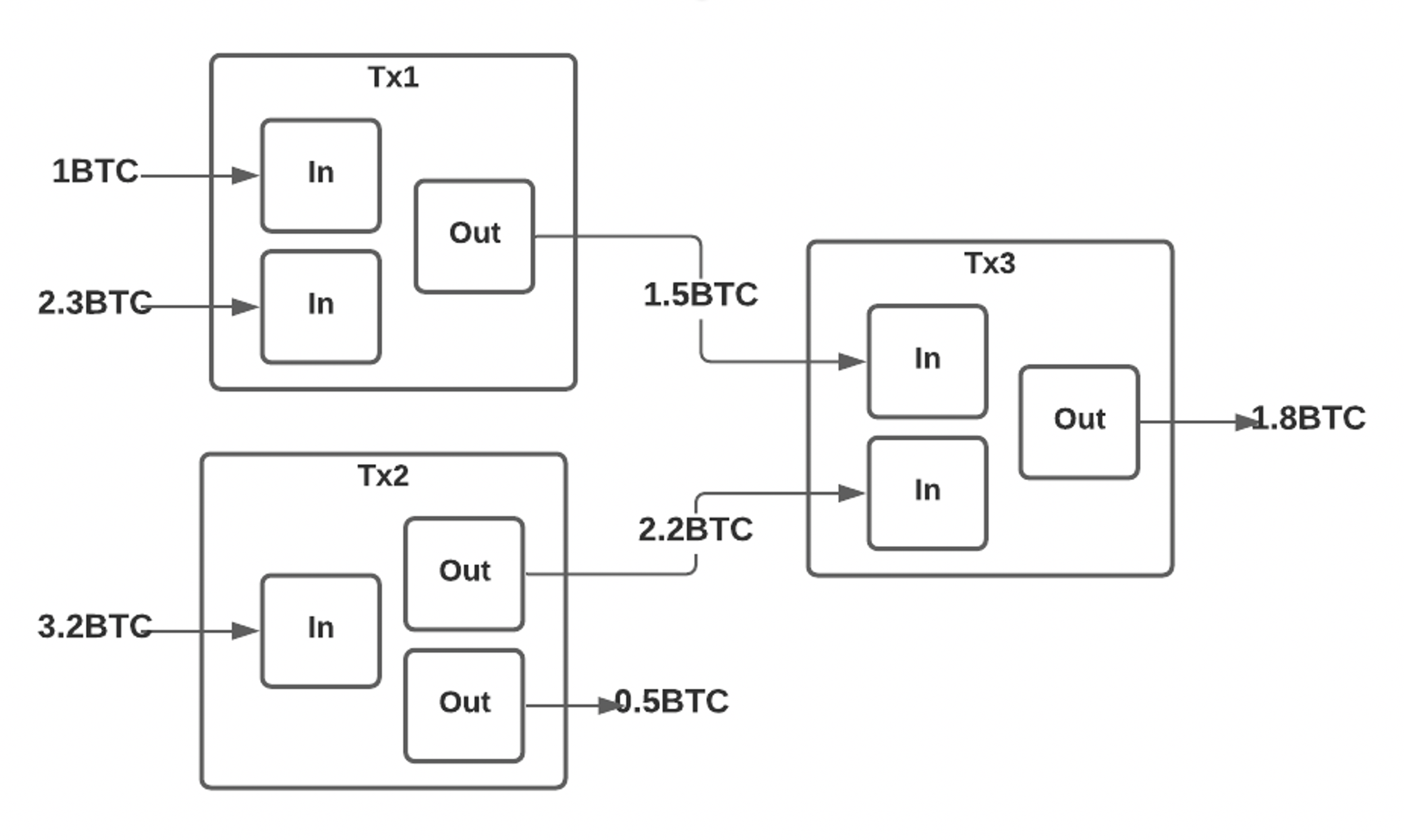}
    \caption{\textit{Concept of UTXOs}}
    \label{fig:utxo}
\end{figure}

There are also conditions for which a transaction in Bitcoin network would be validated and so a successful state transition would then attain; for instance, (1) digital assets involved in the transaction of transfer operations should exist, (2) by enforcing asymmetric cryptography to produce signatures every node should only spend the coins they own and not those of others, and (3) every transaction should be supplied with enough values to the inputs field of every transaction by summing up all the \emph{UTXOs} (Unspent Transaction Outputs) the sending blockchain nodes owned. The concept of \emph{UTXO} is demonstrated in \textit{Fig.~\ref{fig:utxo}}.

\begin{figure}[h]
    \centering
    \captionsetup{justification=centering}
    \includegraphics[width=0.5\textwidth]{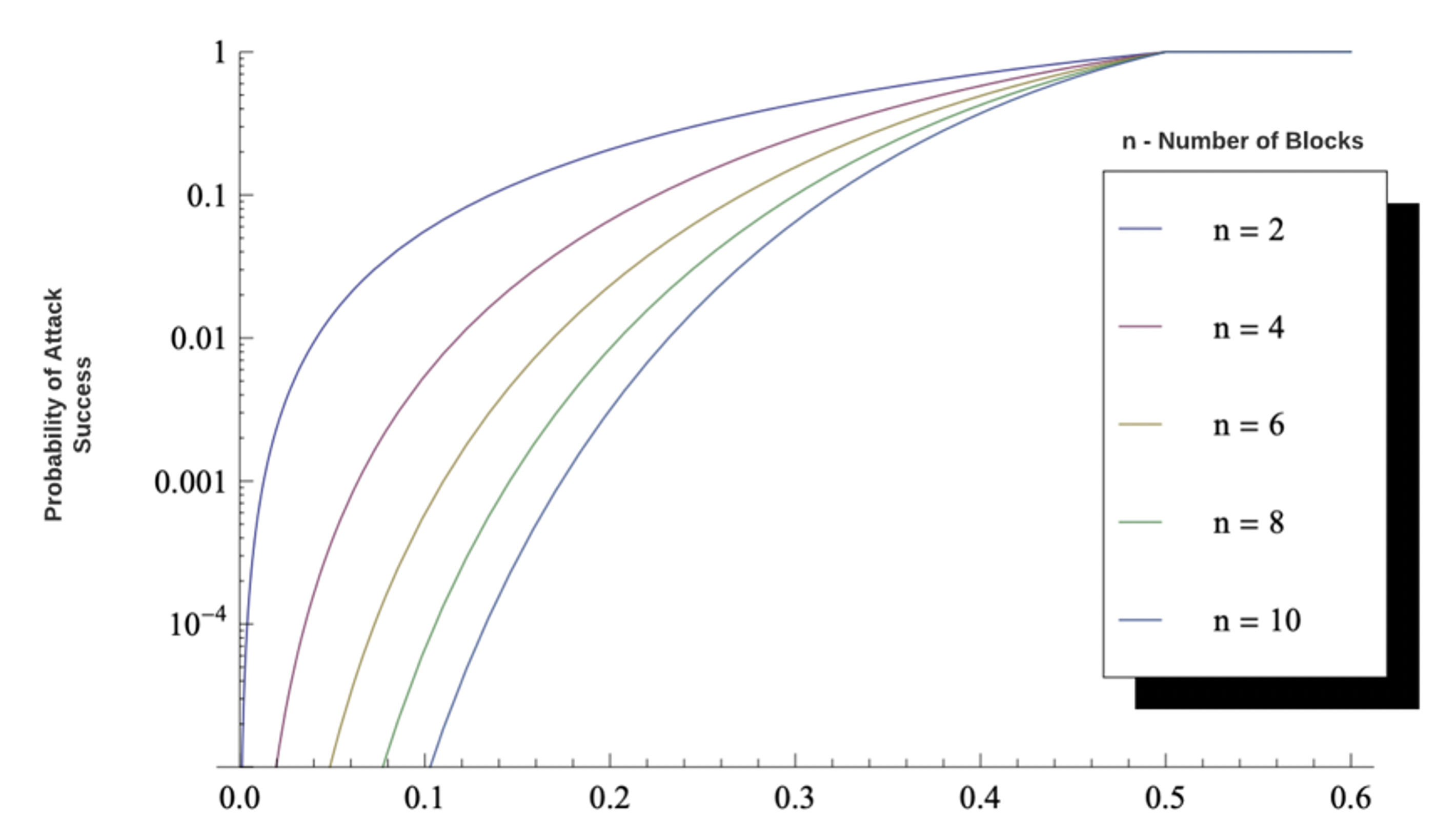}
    \caption{\textit{Analysis of Hashrate-based Double-Spending, Source: M. Rosenfeld}}
    \label{fig:doublespending}
\end{figure}

With the scripting ability of Bitcoin network alongside its Proof-of-Work consensus algorithm requiring validation performed by participating nodes, when the state-transitioned function is validated, the faulty transactions, such as the one sending the same fund twice, will receive an error and therefore be aborted. However, some malicious nodes could try to fork the chain and place a second transaction before the first will require to calculate upcoming blocks with the updated block headers, which would need to create a separate chain longer than the original chain to be the canonical one as nodes are programmed to settle on the chain with largest investment value which is the \emph{canonical chain}. \cite{39} suggested that Bitcoin network could not actually solve the Byzantine Generals problem in general as attackers could theoretically be computationally unlimited and dominate more than 51\% share of the computation power and the overall mining hash rate of the network to perform double-spend operation faster than that on the canonical chain, also known as the 51\% attack under which the analysis on the probability of solving \emph{n} number of blocks consecutively faster than the canonical chain is demonstrated in \textit{Fig.~\ref{fig:doublespending}}. 

There are research and development efforts performed based on Bitcoin model in the field of decentralized electronic payment, such as Litecoin for which these counterparts, any other than the original Bitcoin, were grouped as "\emph{altcoins}", in which some basically are hard-forked versions of Bitcoin while others are having their own underlying native blockchain network with their own consensus protocols, such as Ethereum. With the advent of more and more native blockchain networks with their own consensus protocols and proposed data structures to be supported, blockchain does provide a way for untrusting parties on a peer-to-peer network to agree on contents of a vastly replicated database. The blockchain industry has been focused on exploring more use cases other than merely the decentralized electronic payment using blockchain technology. The development of Blockchain 1.0 has undoubtedly set the premise for new ideas around decentralized autonomous organization and solid basis of the development of Blockchain 2.0 protocols.

\subsection{Overview of Blockchain 2.0 – the Programmable Blockchain}
Given the fact that Bitcoin network offers only basic scripting functionality, with the advent of the open-source Ethereum was published as \cite{33} back in 2014, Ethereum is no longer limited to transaction records, and is more effective and robust than its counterpart Bitcoin. Ethereum blockchain network is a programmable blockchain to perform any arbitrarily complex computation unlike those predefined operations performed in transactions of Bitcoin. Ethereum allows developers to create their own operations of any complexity in smart contracts, utilizing the Turing-completeness programming language and the flexibility brought by the smart contract enabling more possibilities to the blockchain. Ethereum has therefore often been dubbed as web 3.0 due to the fact that the architecture of Ethereum opens up more ideas of general applications with transactions related to data processing and transfer of digital assets, not only the typical use cases, such as decentralized cryptocurrencies.

\subsubsection{States and Accounts}
To get started, Ethereum is an \emph{account-based} blockchain, instead of the \emph{UTXO-based} (Unspent Transaction Outputs) blockchain like Bitcoin network, under which all account states stored locally as a form of state data with the predefined data structure of \emph{Merkle Patricia tree}. As described in \cite{29}, Ethereum blocks contain a list of transactions and the Merkle root hash of entire state tree on transactions which are packed in every block. Every node on the network stores two types of states, namely the overall blockchain state and the state data containing a list of accounts and their associated states.

There are also two general types of Ethereum accounts, (1) \emph{externally owned accounts (EOAs)} with key pairs generated and assigned for signature-based validations or signing transactions on the network, and (2) \emph{contract accounts}, which essentially act as an autonomous agent containing code and would only act once it received a message initiated its functionalities to read from and write to internal storage of the deployed smart contracts. \cite{33} and \cite{37} also proved that contract accounts can send messages to its counterparts with embedded calls to methods of other deployed smart contracts, which is basically another type of account on Ethereum blockchain.

\subsubsection{Smart Contracts and Ethereum Virtual Machine}
\begin{figure}[h]
    \centering
    \captionsetup{justification=centering}
    \includegraphics[width=0.42\textwidth]{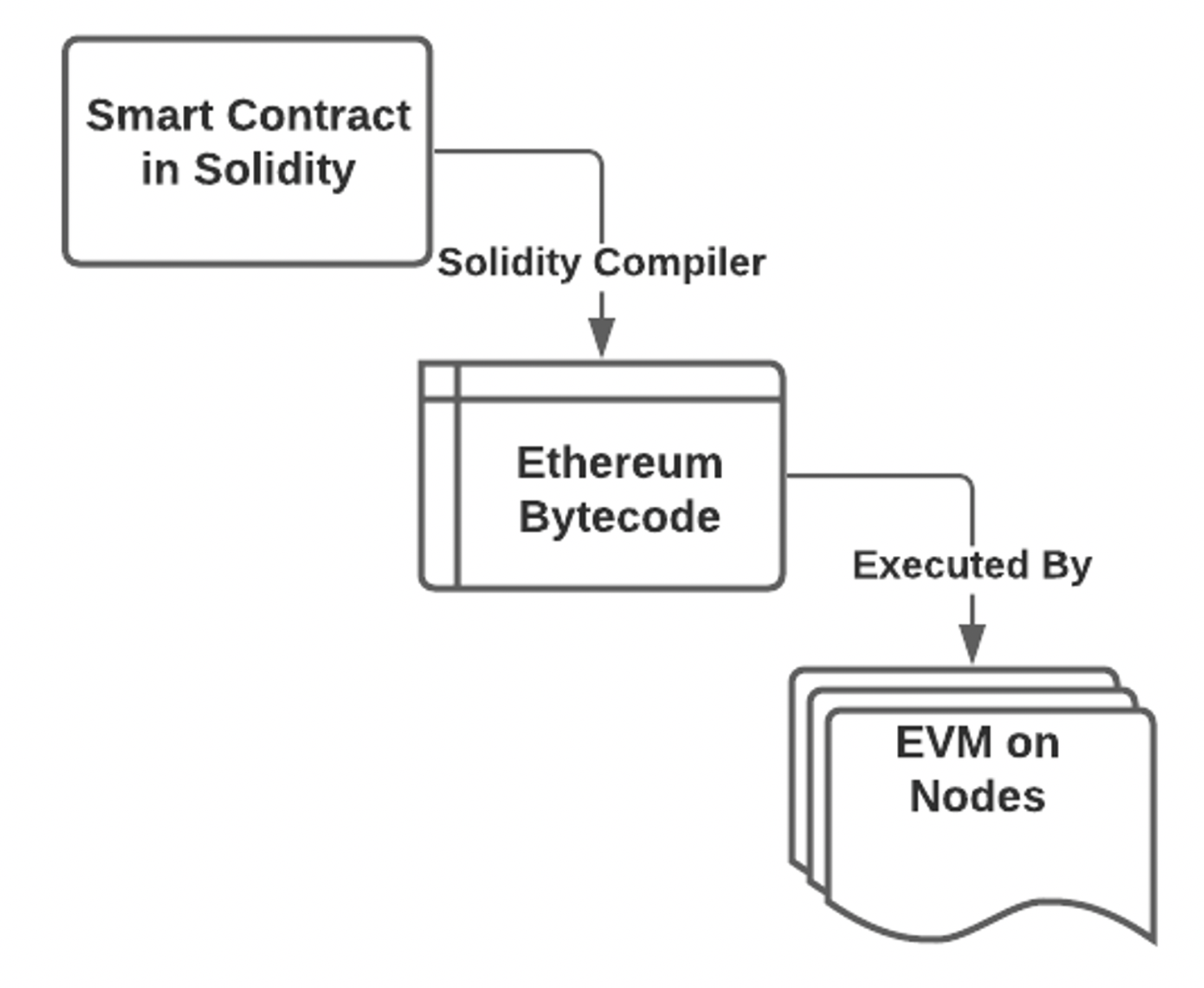}
    \caption{\textit{The Workflow of Ethereum Smart Contract Source}}
    \label{fig:ethereumworkflow}
\end{figure}

Entities on any consensus mode of Ethereum network are able to write smart contracts with methods to define transaction formats, access permission of the methods, state conversion equations suggested in \cite{38} and literally any self-defined rules applied to method declarations with examples demonstrated in \cite{36}. Entities on network could first write and deploy a smart contract to the network for its decentralized applications to interact with, via its dedicated node using the blockchain client, and the smart contract source code, written in Solidity for instance, is then encoded into \emph{Ethereum bytecode}, by the Solidity compiler. The bytecode is added to the data field of a transaction and deployed to the transaction pool of the network queuing to be picked up for further processes. The typical workflow of the smart contract source code is described in \textit{Fig.~\ref{fig:ethereumworkflow}}.

The miners would then pick up the transaction via its node client, pack the validated transaction into a new block and run the hexadecimal bytecode data of the transaction with its Ethereum Virtual Machine (EVM), which is the execution environment for running transaction code to reach a consensus. \cite{33} details a more complex set of instructions to be compiled in the form of smart contracts, to generate operation codes (e.g. \emph{PUSH1 0x60 PUSH1 0x40 MSTORE}) and run based on the operation codes each time a specific method in the smart contract related to a specific transaction is invoked following the rules set within the deployed smart contract. The smart contract could then have state transitioned on each miner's local persistent storage on state data, only if data of the transaction is executed by the EVM successfully. For miner nodes on the network to be able to validate state changes brought by the transaction with the deployed smart contracts, they would be required to run the data, which is the bytecode with operation codes, retrieved from the transaction, on its EVM to check if it is actually a valid state transition. This would impose a large computational redundancy on the network, but it is necessary in order to reach the decentralized consensus.

\subsubsection{Transactions and Gas}
Smart contract of Ethereum is written in Turing-complete languages such as \emph{Solidity} and \emph{Vyper}, allowing for more complex solutions that were not even possible with Script – the deterministic scripting system of Bitcoin network. Making the EVM Turing-complete was a decision made not only for the sake of Turing completeness but also based on the first-class-citizen property in which deployed smart contracts could invoke its counterparts exactly like what the aforementioned \emph{externally owned accounts (EOAs)} could do as defined in \cite{29}; however, such property could also give rise to a potential problem of infinite recursion in which malicious nodes could manipulated and write smart contract methods with transactions causing specific nodes on the network to enter infinite loops and halts, with innumerable gas spent if the block gas limits allow so.

Ethereum has been labelled as Turing-complete based on its guaranteed finality of all executions. Transactions are necessary for state-transitioned methods written in the deployed smart contracts to be invoked. There is also an inherent constraint on the number of computational steps a transaction can perform, which is limited with a concept of "\emph{Gas}" - essentially the cost incurred by totalling all the methods of a deployed smart contract executed in a submitted transaction, based on the gas-spending guideline of each functional pattern predefined in \cite{33}. "\emph{Gas}" is paid in Ether and cannot be avoided in Ethereum main net. The example cost of a transaction is calculated as follows:
\begin{align*} 
Total\ gas\ cost = gas\ spent\ across\ all\ the\ \\ methods\ involved\ *\ gas\ price\
\end{align*}

Nodes will need to specify a maximum amount of gas they are willing to pay per transaction, and if not explicitly stated then the client implementation of the network (e.g. Go-Ethereum, Parity) would determine the amount automatically. If the gas is set too low, it would result in an error leading to a transaction to be reverted, while if the gas limit is set too high, they might ineffectively spend too much gas than it is actually required. The gas price is the amount the transaction sender would be willing to pay per unit of gas spent in a transaction. Higher gas price, which is above the network default average, would result in enhanced chances of getting the transaction to be picked up from the transaction pool and included into a block as the miners tend to pick the highest gas price per unit of gas to get the best return of the computing resources spent per transaction for validation processes performed. The proposed solution should therefore only consider storing data the solution needs, and the storage shall be as leanest as possible as it could cost absurd amount of resources. There are current technologies which could enable decentralized data storage, such as \cite{40}, and therefore decentralized solutions could be built and not required to stick with central database architecture.

Every block, created by miners after the consensus reached amongst nodes running on the network, has a block gas limit similar to the transaction gas limit. Gas limit is the upper bound on amount of computation a transaction can perform on its workflow in the network and that is why people think Ether as the crypto fuels of Ethereum due to a fact that gas cost will be paid in Ether. Another argument why Ethereum is Turing-complete is that if transactions do not have a gas limit specified then attackers could manipulate and expose vulnerabilities of the EVM's Turing completeness property which could likely cause nodes validating the transaction to enter an infinite loop and thus causing a denial-of-service (DoS).

\subsubsection{Block Time and Mining}
Block difficulty could be set to determine the block time so that it would enable for faster block time of exact 12 seconds on Ethereum blockchain. Faster block time implies better performance but decreased security for which multiple block confirmations are expected to secure transactions from double-spending. Blockchain with faster block time would normally have more stale blocks, also known as uncle blocks, due to the increased likelihood of two nodes mining a block simultaneously but only one would be accepted eventually. Independent miners would normally produce stale blocks due to its differing shares of hash power. Therefore, mining pools, like those in Bitcoin network, would always mine a block faster and are more likely to include blocks on the \emph{canonical chain}. \cite{33} introduced \emph{GHOST} (Greedy Heaviest Observed Subtree) protocol allowing miners to get extra rewards if their mined stale blocks are included with any successful block included on the \emph{canonical chain}. This would mean improved security and more profit for independent miners contributing to the security of the network as it becomes more decentralized, compared with the mining pools of Bitcoin network centralizing hash power and resources of the network.

\begin{figure}[h]
    \centering
    \captionsetup{justification=centering}
    \includegraphics[width=0.42\textwidth]{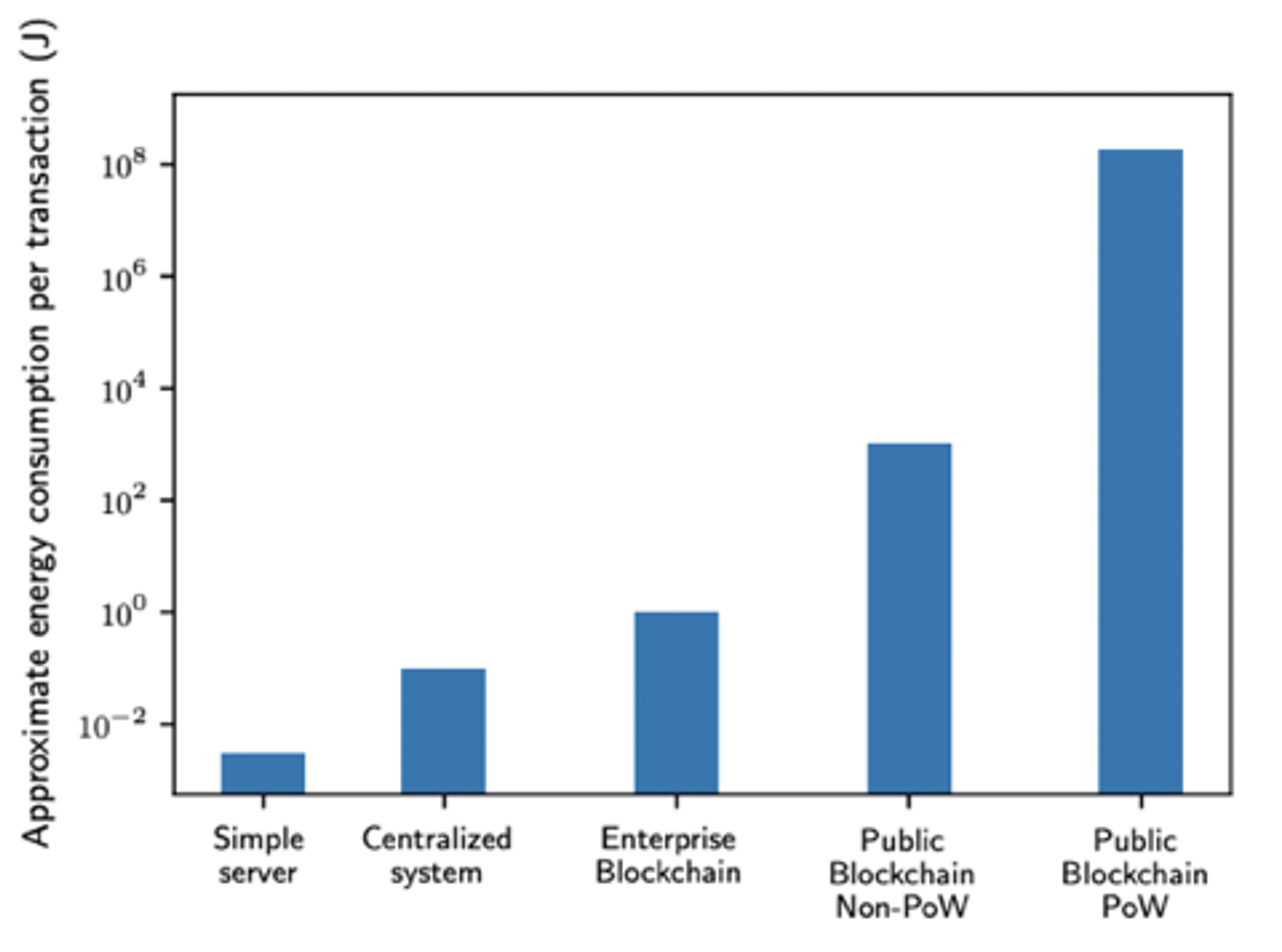}
    \caption{\textit{Energy Consumption of Different Blockchain Implementations, Source: J. Sedlmeir, et al.}}
    \label{fig:energyconsumption}
\end{figure}

Some vulnerabilities, such as the classic \emph{51\% attack}, are caused by the increasing centralization of miners in mining pools. Bitcoin network has proposed to decrease the incentive for miners to join mining pools, which is by far the only way to receive mining rewards from the network, in order to deal with the increasing centralization of the network. Ethereum proposed \emph{Ethash}, which is essentially an \emph{ASIC-resistant} (Application-Specific Integrated Circuit) Proof-of-Work algorithm to tackle the risk of increasing centralization of miners in Bitcoin network, dismissing a memory-hard requirement for which every Bitcoin miner forced using chips that are highly optimized for computing hash values, such as the specified \emph{SHA25} hashing algorithm of Bitcoin mining, to encourage miners using affordable and general GPU instead of relying on ASIC producers, and thus preventing ASIC producers and miners from forging a 51\% share or attack to Ethereum network.
Ethereum is currently in progress to evolve from Proof-of-Work (PoW) to Proof-of-Stake (PoS) protocol also termed as \emph{CASPER} protocol with scaling options available, such as \emph{sharding}, so as to cater to its scalability concerns which exacerbated with innumerable decentralized applications running on Ethereum network, and the infamously high-energy consumption and CO2 emission of mining processes in blockchain networks discussed in \cite{41}. The analysis of energy consumption between different implementation models is visualized in \textit{Fig.~\ref{fig:energyconsumption}}.

The open-source Ethereum blockchain has been planning a major \emph{Ethereum 2.0} (Eth2) upgrade to its network, giving it enhanced scalability and security, as it needs to serve literally any sort of use cases with decentralized applications running on Ethereum blockchain. The first stage of \emph{Eth2}, called \emph{Phase 0} with the beacon chain released, launched in 2020, with later phases such as shard chains for \emph{Phase 1}, shard main net for \emph{Phase 1.5} and fully formed shards for \emph{Phase 2}. \emph{Eth2} would reduce energy consumption, allowing main net to process more transactions with enhanced security. Technically speaking, Ethereum public network will become a Proof-of-Stake blockchain with \emph{shard chains} introduced. This should also bring equally huge benefits on further enhanced security and scalability across different modes of consensus algorithms, such as Proof-of-Authority, available for the development of permission-based enterprise blockchains on Ethereum.

\subsection{Blockchain 1.0 versus Blockchain 2.0 And Later Versions}
Blockchain implementations have been phased based on its development and use cases described in \cite{31,32}. While Blockchain 1.0 with the representative example of Bitcoin network \cite{28} focused on development of cryptocurrencies for peer-to-peer electronic payments, Blockchain 2.0 such as Ethereum \cite{29,33} and Hyperledger \cite{30}, enabled the concept of smart contract, which are more flexible on data structure and functionality enabled by deployed smart contracts. There are also Blockchain 3.0 and Blockchain 4.0 for which the former is focused on developing ÐApps essentially having back-end logic patterns running on a decentralized peer-to-peer platform with a dedicated user interface of it, while the latter is more on matching blockchain technology and its implementations usable especially to those Industry 4.0 business demands, such as process automation, enterprise resource planning and integration of different execution systems respectively.

Starting from Blockchain 1.0, Bitcoin network \cite{28} is indeed the first use case of blockchain technology, aimed at offering a peer-to-peer version of electronic cash. The Proof-of-Work consensus algorithm applied to Bitcoin network, its unique \emph{UTXOs} (Unspent Transaction Outputs) model and the potential hashrate-based double-spending in \cite{35}, which could still exist in Bitcoin network, are still the major characteristics and topics discussed and advanced amongst Bitcoin or even the entire blockchain development community. Blockchain 2.0 moved on enabling programmable blockchains with Ethereum. A variety of underlying concepts of Ethereum blockchain, such as account-based, smart contracts, gas and Ethereum Virtual Machine, as well as explaining how Ethereum is designed and progressed to be different from, and even more capable than, Bitcoin network, are explained in \cite{29,33} as representative examples.

\begin{figure}[h]
    \centering
    \captionsetup{justification=centering}
    \includegraphics[width=0.5\textwidth]{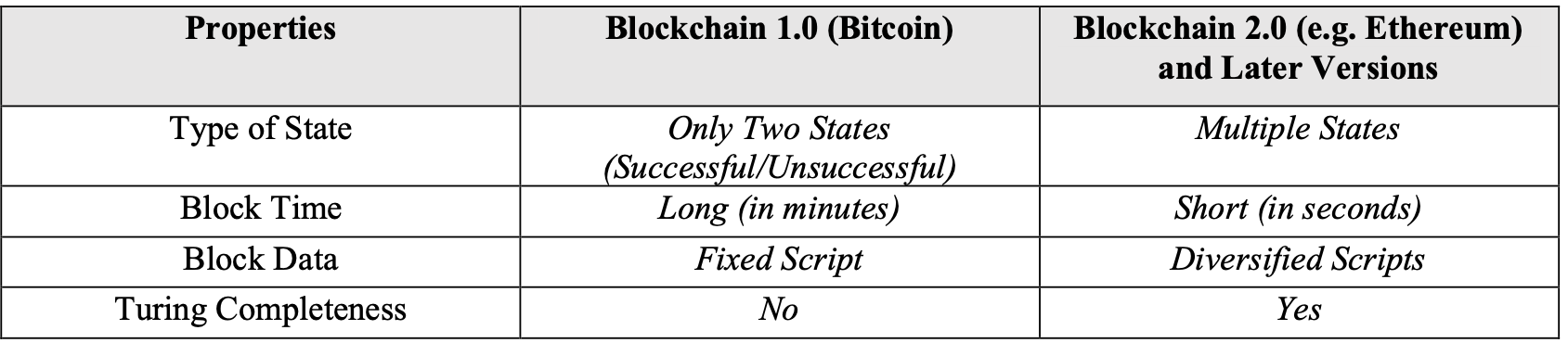}
    \caption{\textit{Comparison of Blockchain 1.0 and Blockchain 2.0 (Incl. Later Versions)}}
    \label{fig:comparisonblockchains}
\end{figure}

There are indeed extensive differences in terms of implementation and usage in both phases. \textit{Fig.~\ref{fig:comparisonblockchains}} demonstrates the most crucial differences in different perspectives, followed by explanations on why modern decentralized solutions would be developed with frameworks and functionalities provided and enabled in Blockchain 2.0.

\begin{itemize}
  \item \emph{State}: In Bitcoin, there are only two states not to mention \emph{UTXO} (Unspent Transaction Outputs) can only have two states as well which could either be "\emph{spent}" or "\emph{unspent}", and so there is no contract or script to keep any internal state other than these two states. Ethereum enables more flexibility to create such contracts by utilizing the concepts of \emph{externally-owned accounts} and \emph{contract accounts}. The multi-state can be defined given the functionalities of smart contracts. Once transactions are validated, packaged into its respective mined blocks and appended to the blockchain, they are no longer allowed to be modified and if such modification on the specific blocks took place, it will invalidate every subsequent block intended to be appended to the blockchain.
  \item \emph{Block Time}: A creation of block in Ethereum main net takes only \emph{12 seconds} (while block time could be specified in enterprise implementations of Ethereum) which is considerably faster than that in Bitcoin network, which would take nearly \emph{10 minutes}. Every transaction will then be validated and packed in a block quicker due to the faster block time in Ethereum but it would lead to decreased security, attributed by the faster block time, which has already been addressed by multiple block confirmations.
  \item \emph{Storage}: In Bitcoin network, only fixed scripts and data can be stored in a block while self-defined scripts, in a form of smart contracts, can be executed with states stored on Ethereum blockchain implying that more methods are enabled and therefore many different applications can be implemented on Ethereum or its Blockchain 2.0 counterparts.
  \item \emph{Turing Completeness}: The script in Bitcoin network is designed not supporting loops so infinite loops can be prevented, while Ethereum provides more flexibility in script writing of its smart contract implementations and Turing completeness as it employs different methods to eradicate infinite loops.
\end{itemize}

While the open-source Ethereum blockchain has been planning a major \emph{Ethereum 2.0} (Eth2) upgrade to its network to address scalability concerns, there has been an array of blockchain frameworks in Blockchain 2.0, are available for developing decentralized solutions based on a concept of enterprise blockchain, such as Hyperledger Fabric depicted in \cite{30} or Tendermint Core developed based on \cite{42,43}. Like Ethereum, all of these have been seeking to prove that enhanced security, enhanced degree of decentralization and enhanced scalability are not at odds.

\subsection{Related Work of Blockchain Implementations}
With the advancement of blockchain development in recent years, there have been some existing blockchain innovations developed in different domains and in combination with other emerging technologies to decentralize software systems with centralized architecture of different purposes. A blockchain-based digital certificate system and blockchain-enabled system for personal data protection were proposed in \cite{49,50} respectively. \cite{44,45} have also given an overview of blockchain-based applications developed in different domains where a variety of examples of blockchain systems and use cases are developed, could be found in healthcare domain as depicted in \cite{46,47} focused on decentralizing health record management and storage.

Blockchain innovations have also been implemented across the supply chain industry, and some are specifically with use cases of decentralizing and improving product traceability and anti-counterfeiting aspects of the supply chain industry. \cite{55} has proposed a concept of blockchain system to enhance transparency, traceability and process integrity of manufacturing supply chains, while an Ethereum-based fully-decentralized traceability system for Agri-food supply chain management named AgriBlockIoT was developed in \cite{56}. Furthermore, a novel blockchain-based product ownership management system, a blockchain-based anti-counterfeiting system coupled with chemical signature for additive manufacturing and ontology-driven blockchain design for supply chain provenance, are also detailed in \cite{57,58,59} respectively.

Some implementations are conceptualized and developed based on computation-extensive permission-less blockchain networks and consensus algorithms, aiming at full decentralization over scalability and stability of such decentralized systems developed. Instead of developing blockchain implementations based on conceptual design, decentralizing legacy anti-counterfeiting systems with centralized architecture already implemented in the industry, further with blockchain innovations integrated with, would be a more pragmatic way to start with so as to provide timely support to improve the snowballing situation of product counterfeits in supply chain industry.

\section{The NFC-Enabled Anti-Counterfeiting System - NAS}
\begin{figure*}[h]
    \centering
    \captionsetup{justification=centering}
    \includegraphics[width=0.9\textwidth]{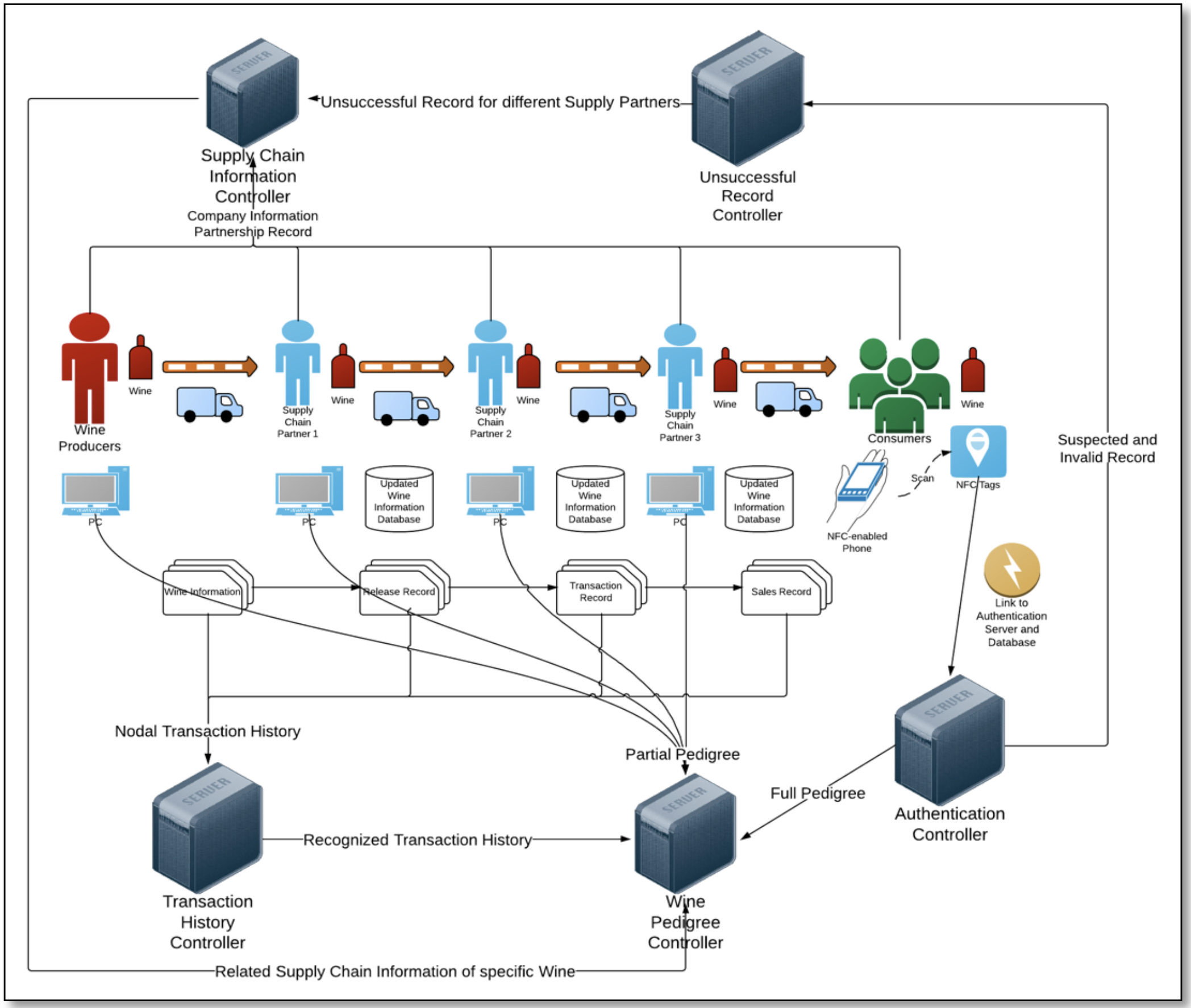}
    \caption{\textit{The System Architecture of NAS, Source: Neo C.K.,Yiu}}
    \label{fig:nasarchitecture}
\end{figure*}

One of the solutions to answer the growing concerns on product counterfeiting in different supply chain systems of wine industry, is the \emph{NFC-enabled Anti-Counterfeiting System (NAS)}. \cite{25} was developed and implemented back in 2014, aiming at providing an innovative and fully functional alternative, based on \emph{Near-field communication technology} and cloud-based microservices architecture with centralized storage structure, solely hosted by any winemaker, to help improve the worsening situation of product counterfeiting especially for the wine industry. NAS with centralised data architectures, which are predominantly based on typical and familiar cloud-based client-server architecture style, is demonstrated in \textit{Fig.~\ref{fig:nasarchitecture}}.

The whole NAS solution is consisted of FIVE main components, which are (1) the back-end system with web-based database management user interface for wine data management performed by winemakers, for which management of data columns of specific wine products which are in their custody can be performed by winemakers, (2) an NFC-enabled mobile application, \emph{ScanWINE}, for tag-reading purpose of wine products at retailer points for wine consumers or supply chain participants of the supply chain before accepting these wine products, (3) another NFC-enabled mobile application, \emph{TagWINE}, performing tag-writing purpose for wine at wine-bottling stage by winemakers, (4) the NFC tags packaged on wine bottlenecks for those purposes and actions, and (5) any NFC-enabled smartphones or tablets of supply chain participants and wine consumers along the supply chain.

There are FOUR major categories of data to be processed along the supply chain with NAS, namely the (1) nodal transaction history data, (2) supply chain data, (3) wine pedigree data which is processed with its dedicated controllers based on their predefined schema and data models, and (4) unsuccessful validation data returned from any unsuccessful validation at the stage of accepting wine products. As wine products being processed and handled by different nodes along the supply chain with data updated by scanning NFC tags of the wine products using tag-reading \emph{ScanWINE} with the state of wine record to be updated accordingly to the database. These categories of data are updated along the supply chain until the point of purchase at which wine consumers use tag-reading \emph{ScanWINE} to scan NFC tags and retrieve data such as the wine pedigree data and transaction data for real-time validation to determine if a wine product is counterfeit or not.

A unique identifier is assigned to each wine product and is written into the NFC tag. Such tag-attached wine products are then shipped from winemakers to different nodes along the supply chain. During the transportation process of these wine products along the supply chain, each involved node could scan the NFC tags and adds the aforementioned four categories of data into these NFC tags respectively. In this way, the next node can check whether or not the wine products have already passed through the legitimate supply chain. If any inconsistency is found at any node, such wine products may be considered as counterfeits and should be returned to winemakers. However, once the wine product reached post-purchase stage and circulated in any customer-to-customer markets, its authenticity is no longer guaranteed, as anyone who has an NFC reader can interfere and clone tags' data. Therefore, it is important to develop anti-counterfeiting systems that could work even when the data stored in tags is cloned in post-purchase supply chain with attacks detected and prevented on any potential adversary changes.

\section{Security Analysis on NAS}
In this chapter, a series of security analyses are performed on NAS, with findings of these security analyses also elaborated and discussed so as to identify which areas of NAS could be improved and strengthened, in terms of security, with the conception of decentralized product anti-counterfeiting and traceability solution enabled by blockchain technology.  

\subsection{Asset Analysis}
The asset analysis of NAS, as described in \textit{Appendix~\ref{a1}}, lists out ten constituent assets of NAS which are categorized into components of hardware, software and data. Hardware assets are NFC-enabled mobile devices and NFC tags, while software assets of NAS are the two NFC-enabled mobile applications and the database operation web application. As described in the system overview of NAS, the data model of NAS is indeed based on four types of data assets, namely data of, wine pedigree, transaction history, supply chain and unsuccessful records.

The (1) \emph{NFC tags (A04)}, (2) \emph{transaction history data (A06)} of a wine record and the (3) \emph{backend database (A10)}, which is solely managed by winemakers, storing all sorts of data components listed in the asset analysis, are amongst the three most probable system components susceptible to security risks according to ratings assigned to each component of confidentiality, integrity and availability (The CIA methodology).

\subsection{Threat Analysis}
Based on the result of asset analysis and system components identified, which are susceptible to security risks, a threat analysis is therefore performed against NAS, with every threat as described in \textit{Appendix~\ref{a2}}, for which every possible threat to NAS could be categorized into either (1) physical NFC tag threats or (2) system threats.

Regarding \underline{\emph{physical NFC tag threats}}, the most probable and risky threats identified are: 
\begin{enumerate}
  \item \emph{Tag cloning (T01)} for which each NFC tag used for product tagging and anti-counterfeiting purpose has a unique identifier. If the identifier information is exposed to attackers, data stored in a tag could easily be cloned into another tag.
  \item \emph{Tag disabling (T03)} for which adversaries could take advantage of wireless nature of NFC systems in order to disable tags, from any further interaction with NFC tags, temporarily or permanently by changing the state of NFC tags.
  \item \emph{Tag's data modification (T04)} for which NFC tags use writable memory and so an adversary can take advantage of such feature to modify or delete valuable data from memory of any involved NFC tag, during any tag-reading and tag-writing process. Unsecured configuration or even misconfiguration on NFC tags could also allow attackers compromising NAS as a whole.
\end{enumerate}

These physical NFC tag threats are primarily attributed by weak and fully-centralized authentication and authorization mechanism adopted to change data states stored in NFC tags and any unsecured configuration on NFC tags during every tag-writing and tag-reading process.

While regarding \underline{\emph{system threats}}, the most probable and risky threats identified are:
\begin{enumerate}
  \item \emph{Man-in-the-middle relay attack (T08)} for which in a relay attack, an adversary acts as a man-in-the-middle. An adversarial device is placed surreptitiously between a legitimate NFC tag and mobile applications or mobile applications with dedicated backend database which is on logged-in state. Adversaries could obtain unauthorized access or unintentional information disclosure on the confidential data related to transactions or the supply chain as a whole.
  \item \emph{Tracking and tracing (T10)} for which via sending queries and obtaining same responses from an NFC tag at various locations, it can then determine where an NFC tag of a specific product is located physically with location data supplied. A malicious hacker may also obtain login data, via brute-forcing or dictionary attacks, to spoof and login to mobile applications or web applications registered with the same login data as legitimate users.
  \item \emph{Denial-of-service (T11)} for which DoS attacks are usually physical attacks, such as jamming the system with noise interference, blocking radio signals, removing or even disabling NFC tags, causing different system components or the entire system to work improperly. Without sufficient auditing logs and monitoring functionalities of different system components, it would be unable to investigate into any system misuse and compromise, security breach over leakage of confidential data.
  \item \emph{Spoofing attack on data of product records (T14)} when attackers get some information about the identity of NFC tags either by detecting communication between mobile applications and legitimate NFC tags or by physical exploration on these NFC tags, attackers could then clone the NFC tags. Poor code quality on microservices enabling interaction between mobile applications and the backend database of product records, such as weak authorization and authentication with dummy passwords or without database backup path, could lead to data theft on the confidential data of processed product records maintained in NAS.
\end{enumerate}

These system threats are majorly attributed by the single-point processing, storage and failure due to the fact that operations and data of NAS are managed and controlled solely by winemakers as the anti-counterfeiting and traceability features of NAS are built on specific winemakers and around industrial operations enabled by NFC technology or other tag-communication technologies. Furthermore, vulnerabilities such as weak authentication and authorization as well as in lack of sufficient auditing logs and effective API monitoring tools could also give rise to threats, such as man-in-the-middle relay attack, tracking-and-tracing and spoofing attack, for which adversaries could manipulate vulnerabilities to obtain unauthorized access or unintentional information disclosure over the confidential data such as the transaction data, wine pedigree data or supply chain data. The disclosure of confidential data would then lead to adverse manipulation on product records and so disability of anti-counterfeiting functionalities of NAS could be expected. Adversaries could also make use of vulnerabilities, such as unsecured configuration on servers, poor security implementation over the code base on possible attacks as well as lacking audit logs and API performance monitoring, to perform denial-of-service (DoS) attack on different system components of NAS affecting its service availability, stability and performance.

\section{Discussion of Research Results}
With blockchain fundamentals explained, how blockchain technology could impose security upgrades to existing product anti-counterfeiting and traceability systems, such as NAS, of supply chain industry, and the results gathered from security analyses performed on NAS, this chapter will cover the summary of vulnerabilities identified in existing product anti-counterfeiting and traceability systems, opportunities of decentralizing anti-counterfeiting and traceability in supply chain industry and potential concerns on developing decentralized solutions for supply chain industry.

\subsection{Summary of Vulnerabilities on Centralized System Architecture}
Amongst NAS and other existing anti-counterfeiting alternatives with centralized architecture, utilizing wireless tag communication technologies, there could be at least three common probable counterfeit attacks applied to these anti-counterfeiting solutions via manipulating threats listed under the \emph{physical NFC tag threats} and \emph{system threats} according to the threat analyses performed: (1) modification of product records stored in tags, such as fabricating product identifiers or vintages of any product, (2) cloning of metadata stored in tags such as those genuine product records to any counterfeit product tag, and (3) removal of a legitimate tag from a genuine product and its reapplication to any other counterfeit products.

It has come to a point that even though the implementation of NAS itself is already more effective and secured than most of its typical supply chain anti-counterfeiting and traceability counterparts, with original product records being validated at any node along the supply chain. The centralized architecture of NAS could still pose risks in data integrity and product authenticity as any node, not only winemakers, along the supply chain have full control of product records stored in their own database architectures. In case different nodes along the supply chain are untrusting to each other, there could still be possibilities that a product record being duplicated adversely leading to a situation that product consumers could still purchase a product counterfeit at retail points, with fabricated wine records retrieved from NAS or its counterparts implemented in specific supply chain industries.

The typical architecture of centralized supply chains creates several concerns. First, there is a tremendous processing burden on servers, since significant numbers of products processed by multiple supply chain nodes. Second, substantial storage is required to store authentication records for every single processed product. Third, with centralized systems, traditional supply chains inherently have the problem of single-point failure and so potential service downtime and data loss could be expected. All in all, centralized product anti-counterfeiting and traceability systems, such as NAS, rely on a centralized authority to combat counterfeit products which would result in \emph{single-point processing, storage, and failure} and those potential attacks via manipulating the security threats identified in threat analysis performed against NAS.

\subsection{Opportunities of Decentralizing Supply Chain Anti-Counterfeiting and Traceability}
\begin{figure}[h]
    \centering
    \captionsetup{justification=centering}
    \includegraphics[width=0.5\textwidth]{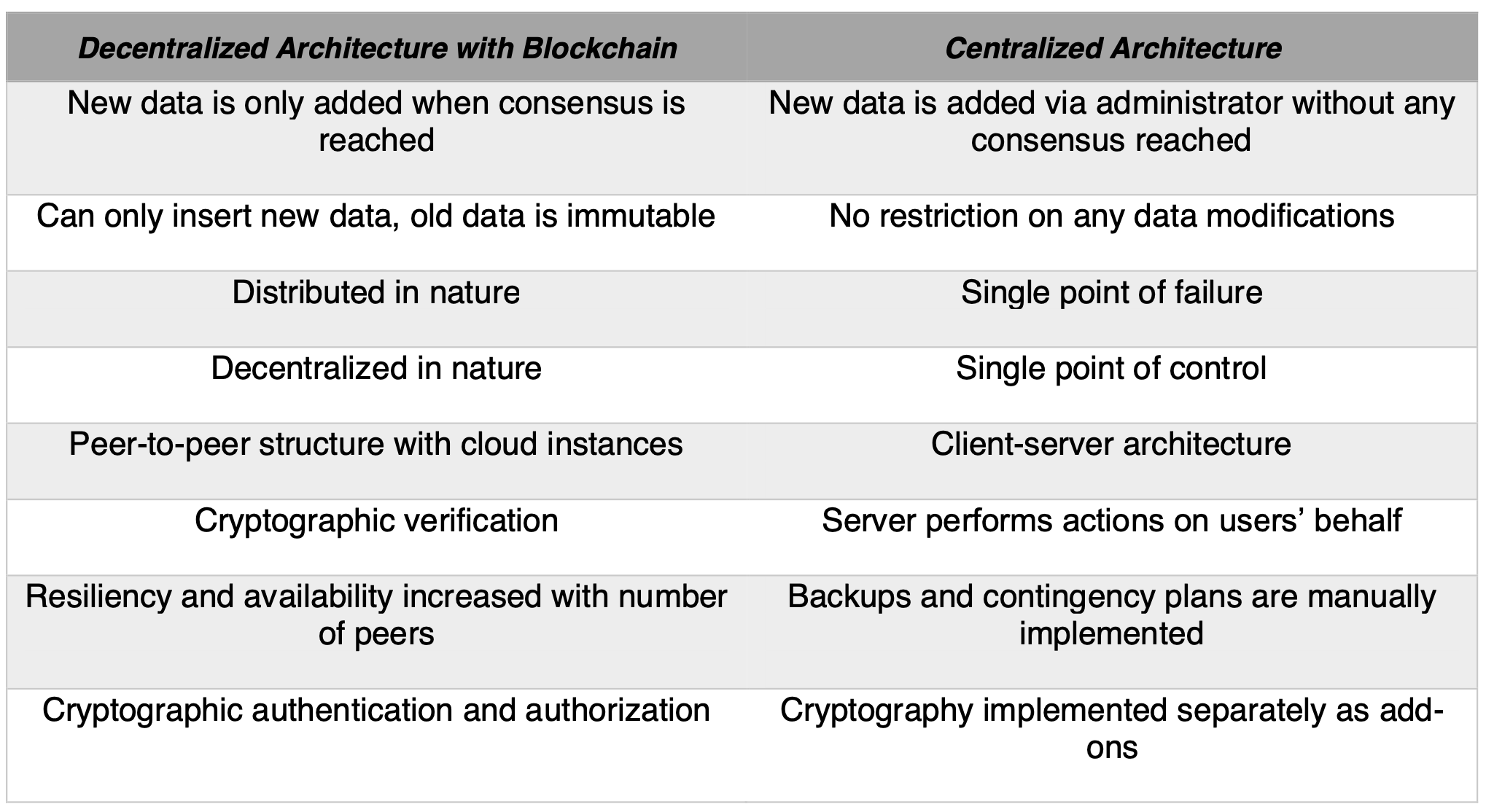}
    \caption{\textit{Comparison of Decentralized and Centralized Architecture}}
    \label{fig:comparisontable}
\end{figure}

To better prevent risks and overcome threats with vulnerabilities initiated by centralized architecture, Blockchain Technology (or other Distributed Ledger Technologies built with decentralized networks) stands out as a potential framework to establish a modernized, decentralized, trustworthy, accountable, transparent, and secured supply chain innovation against counterfeiting attacks, compared with those developed on centralized architecture, with comparison between two as detailed in \textit{Fig.~\ref{fig:comparisontable}}.

Given a variety of advantages, such as prevention of single-point failure, better resilience and availability of being applied amongst supply chain participants, introduced with the blockchain technology and concept of decentralized application, to have a more secured and sophisticated supply chain system against counterfeiting attacks, it has well proven that decentralized supply chain anti-counterfeiting and traceability are in demand and worth developing and implementing in supply chain industry, starting with a new solution or developing a novel prototype with a decentralized architecture, based on legacy solutions, such as NAS, to reinforce the innovative idea of product anti-counterfeiting. There are also a variety of opportunities that autonomous and decentralized supply chain anti-counterfeiting and traceability solutions could bring to supply chain industry as explained in the following.

\subsubsection{Improved Data Integrity of Supply Chain}
With the advent of blockchain technology and other technologies such as distributed file storage. A multi-layered data storage and validation mechanism, involved with on-chain and off-chain data operations, on product records could be implemented in decentralized solutions of supply chain anti-counterfeiting and traceability.

The on-chain and off-chain storage and validation could then be in place to ensure data integrity and to prevent the decentralized solutions from any attempted attacks, such as cloning attacks on NFC tags, modification attacks in case product identifiers and signatures stored in NFC tags are inconsistent with its counterparts stored in the backend databases or on-chain storage with deployed smart contracts, or reapplication attacks in case both read count and write count are inconsistent to its counterparts stored off-chain and on-chain respectively. 

With the decentralization enabled by the blockchain network, data integrity of processed product records is further improved which could further coupled with a concept of digital asset tokenization representing every single product processed on the decentralized solutions, owing to the fact that any state change on specific product record could only be initiated with invoking methods on an open smart contract deployed to the network, with its transaction now needing to be validated with consensus reached, with other nodes held and run by different participating entities of supply chain industry, on the network. The immutability of transaction states related to any state transition of specific product record operations would mean that any state change processed on the network could be referred and queried based on individual transaction hashes and block numbers, which could further be confirmed on blockchain explorers connected to blockchain nodes running on the specific blockchain network.

\subsubsection{Strengthened Security Considerations}
Given the security considerations deployed to different possible validation mechanisms of any product record validation operation to improve data integrity, and according to the findings from the threat analysis performed against NAS, a variety of security attacks existed in NAS are no longer valid and could well be prevented with errors thrown if those attacks are detected before a state transition could be completed on a specific product record.

Any state transition on product records is required on-chain and off-chain validations performed against the data of product records, with respective transactions also validated by other blockchain nodes running on the same blockchain network. These validation steps of product record validation operation are now required to include signature generation and signing procedures, with key management modules offered to users, on any attempted state transition on product records. Regarding security considerations applied to the deployed smart contracts which will be required if any blockchain 2.0 implementation is adopted to decentralize the supply chain anti-counterfeiting and traceability, multiple validation syntax on specific conditions are developed and included in different methods of smart contracts to prevent from being manipulated by potential attacks, such as reapplication attacks in which the on-chain write count and its counterpart stored off-chain do not actually match. 

Design patterns with role-restriction concept of the deployed smart contract could also be introduced, so as to enable access authorization for authenticated node accounts held by specific entities of supply chain industry, to different methods defined in the deployed smart contracts. The security model on data integrity of NAS is currently based only on operations before the point of purchase, and it could further be extended when the supply chain anti-counterfeiting and traceability functions are properly decentralized, as long as the post-purchase wine consumers of consumer-to-consumer market are also registered entities or even running nodes in the decentralized solutions.

\subsubsection{High Availability of System Functionalities}
Ensuring high-level operational performance of different system components to maintain system functionalities for different product record operations is key to the system implementation of NAS or any other supply chain anti-counterfeiting and traceability systems. With opportunities of system security and data integrity on product record now highly dependent on the system decentralization, availability of data and states stored, as well as those decentralized system components become more significant to the overall availability of system functionalities.

Regarding the decentralized solutions, availability and resilience on data and states, stored on blockchain network or any other decentralized system components, are assured and even be enhanced with increasing number of blockchain nodes running on the blockchain network owing to the fact that each node of these networks keeps the copy of the states stored in persistent volume dedicated to these distributed nodes. Availability of the blockchain network would also be enhanced with increased amount of blockchain nodes running on the blockchain network in which availability could be preserved as long as there is at least a blockchain node running on the network.

Dedicated persistent volume storage could also be assigned to each node instance running on the blockchain network to store blockchain states and individual chain data, so as to benefit from faster synchronization and data recovery on states to any new blockchain nodes connected to the network. This will then assure the availability of blockchain nodes and the blockchain network as a whole, as failed blockchain nodes could be reconnected to synchronize and process transactions sent to the network immediately. Nonetheless, availability of the data stored on-chain will be preserved with the smart contracts deployed to the blockchain network as long as it is actively running and mining new blocks constantly. The off-chain database and the app-backend service could also be made distributed with individual instances with which every participant under the decentralized solutions could now host their own instance of the supply chain anti-counterfeiting and traceability ecosystems.

\subsection{Potential Concerns on Development of Decentralized Solutions}
According to the summary of vulnerabilities on centralized system architecture, with opportunities of developing the decentralized solutions also identified, a set of fundamental system requirements of a decentralized version of NAS, namely the Decentralized NFC-Enabled Anti-Counterfeiting System (dNAS), is also proposed with potential concerns on developing such decentralized solutions elaborated in the following.

\subsubsection{Manageable System Integration Model}
It is understood that not every user of the decentralized solutions would possess with in-house technical capacity to maintain its own instance of system components constituting the decentralized solutions. A manageable system integration model will need to be in place to help promote adoptions of the decentralized solutions from its centralized counterparts, which have been implementing and adopted in the wider supply chain industry, amongst different potential users in supply chain industry and for different stakeholders of industries to collaborate for good to help improve the worsening situation of product counterfeits, with the process integrity also conserved.

Such manageable system integration model could provide another layer of indirection, allowing potential users to safely manage their own keys, the backup of product records in case anything unexpected goes wrong with system components of the decentralized solutions, while an organization or alliance of major industry participants could act or be voted as leaders and host system components such as microservices, mobile applications and even the decentralized blockchain nodes forming a blockchain network, as a fail-over and manage requests from these potential users of the decentralized solutions. Common data model of product records, applied to the decentralized solutions, should also be defined with additional metadata added after completing different types of industrial operations or steps declared in product data operation such as data validation and data creation steps both off-chain and on-chain, in order to facilitate a seamless process of data migration and integration.

\subsubsection{Degree of Decentralization}
The degree of decentralization is dependent and based on the chosen mechanism of manageable system integration, which also requires promoting adoption of such decentralized solutions implying that some software components of the decentralized solutions are still expected to be hosted by intermediaries such as backend database, mobile applications and backend application, with users only keep hold of secrets or instances of decentralized system components such as blockchain nodes assigned to registered entities to the decentralized solutions.  

The decentralized model could be substantiated with the blockchain network and its chosen consensus algorithm. Implementations of a blockchain network with its dedicated blockchain interface do provide certain degree of decentralization when it comes to transactions being validated and packed in a block on-chain with the respective methods of deployed smart contract invoked as well as proposing state changes on-chain registry for peer authentication and the blockchain network. Decentralization around smart contract management and deployment, management of software component instances of the decentralized solutions, and management of the distributed network protocols, might be limited depending on the choice of blockchain implementations adopted but would definitely not be solely managed by product producers in the supply chain industry.

\subsubsection{Limited Scalability}
Scalability can somehow contradict with the degree of decentralization defined in the current setting of blockchain implementations, for which decentralized solutions generally take longer computation time for the same set of operations performed with centralized solutions, such as NAS, and so it is expected that decentralized systems are generally less scalable than its centralized counterparts owing to the fact that consensus is needed for every state changed on the data of processed product records.

The multi-layered validation and creation mechanism of product records processed in the decentralized solutions could in all likelihood imply decreased scalability as they could now involve more steps to store or even update representations of any product record, given these state transitions on product records involved are required to be processed in the decentralized system components as well. In case the number and size of product records growing with more products circulating in the supply chain, longer computation time is definitely needed for these product record operations with more computation resources committed on data processing of product records stored in the off-chain backend database structure. Data stored and processed on decentralized system components, such as the blockchain network, should be kept minimum, given the computation resources needed for these decentralized processes could be exponential compared with its centralized counterparts. The requests sent to the existing microservices and blockchain interface could be handled sequentially for which a new request could be processed only if the previous one is completed. The single-threaded handling of these microservices, involved in any end-to-end product record operations, could possibly hinder the scalability of the decentralized solutions as a whole.

Given the extra decentralized processes required in the decentralized solutions from the point of invoking endpoints made available in blockchain interface all the way to the blockchain network. The size of blockchain will grow bigger when the time goes on with more transactions validated and packed in blocks mined, not to mention a new block could be created in an interval of block time defined when initiating the blockchain network. It could take fairly long period of time for any new blockchain node, to synchronize with other blockchain nodes to get to the latest global states of the network. The long synchronization time would hinder the user experience for participants using the decentralized solutions when the size of blockchain is too bulk. Though there are potential scalability concerns on any proposed decentralized solutions, it is still worth decentralizing supply chain anti-counterfeiting and traceability, if comparing with the benefits brought by decentralization, such as the strengthened data integrity and improved system security with distributed instances enabling individual nodes along the supply chain collaborating to combat product counterfeits.

\subsubsection{Potential Security Vulnerabilities}
Every registered instance of the decentralized solutions, is normally assigned with a blockchain node and an account of which a key pair could be generated and assigned to registered instances for their storage and management. Key pairs are required to validate and send transactions with its local blockchain node via the chosen blockchain client protocol. The same key pair could be retrieved and utilized over and over again without a concept of proper key rotation and management, and it is possible that the key pair could be compromised and hence the aforementioned attacks could still be made possible to create vulnerabilities and threats to the decentralized solutions.

Following the compromised key secrets, \emph{distributed denial of service} (DDoS) could also be made possible as long as the blockchain node is hosted with enough computation resource, and it is possible to spam the blockchain network with huge amount of transactions to be processed. The denial-of-service attack could also be performed by any malicious registered node though they are part of the supply chain industry. A distributed key management module with key rotation functionalities to store and manage the key secrets should be implemented in the decentralized solutions, and a security authentication layer should also be added on top of the key management module whenever the blockchain interface, owned by the same registered instance, retrieves the key pair.

While for the data of product records stored and processed on decentralized components of decentralized solutions, such as the smart contracts deployed to the blockchain network and the service instance of any distributed file storage implemented, any blockchain node assigned to the registered entities could interpret and retrieve product record subsets if required if the unique identifier of processed products is supplied, as these product record subsets stored with the deployed smart contracts might not be encrypted and obfuscated with any hash algorithms. Though application of data security is really dependent on which types of blockchain network and consensus algorithm are adopted, which could generally categorized either into "\emph{permissioned}" or "\emph{permissionless}", the decentralized solutions could adopt. Any data stored and processed on any decentralized system components should be kept minimum, obfuscated and even encrypted to prevent from any potential malicious manipulation. 

The publicity of the smart contract source code could also cause security vulnerabilities. Unlike the source code of different system components in NAS which has the option to have its code base open-sourced or completely privatized, smart contract code of decentralized solutions is always open and easily accessible by blockchain nodes running on the same network, and so malicious users, of the decentralized solutions, running blockchain nodes could look for human-induced vulnerabilities if any method of the deployed smart contracts is not implemented correctly.

\subsubsection{Privacy Concern}
As discussed in the potential security vulnerabilities, lacking key rotation mechanism would imply the same public address is possible to be mapped to an actual registered node with which the system role identifier could further be mapped to a true identity of the representative organization or entity of the supply chain industry, by other registered nodes running on the same blockchain network if the decentralized solutions are developed in a setting with \emph{permissioned blockchain implementations}.

Although public addresses stored on-chain could be obfuscated with hash functions applied, events could be emitted when methods of the deployed smart contracts are invoked, whenever there is a new transaction initiated on product record operations related to the same public addresses. The related events are later received by the event listener of every blockchain service instance. With more events emitted involving the same set of public addresses, it is more likely a specific public address could be mapped to an actual registered instance, and so its transaction volume could still be derived by other users of the decentralized solutions, which could potentially be its competitors. 

In addition to public addresses, these data fields could directly relate to physical entities, and cause privacy concern if there is no privacy-preserving technology in place for these sensitive data fields. If the respectively NFC tags are not deactivated properly when the respective products are consumed or transferred, it could possibly lead to a privacy threat based on any unencrypted or unobfuscated data field of specific product records stored in the NFC tags. Privacy-preserving technologies are required with use cases defined, based on chosen mechanisms on data processing and validation procedures to be included in the decentralized solutions.

\section{Conclusions}
With the current challenges on product counterfeiting of supply chain industry, the research is performed basing on a set of research objectives with a main research question and three sub-questions addressed. The contributions of this research are to determine if existing anti-counterfeiting and traceability solutions of supply chain industry could benefit from a certain degree of decentralization applied to those existing solutions already implemented in the industry, via a series of security analyses, functional analyses on possible opportunities with decentralizing supply chain anti-counterfeiting and traceability, and to define a list of fundamental system requirement of a decentralized version on NAS, which is one of the software solutions being currently implemented in supply chain industry.

A thorough blockchain fundamental explanation is covered in the research efforts, ranging from a core overview and characteristics of blockchain technology, the Blockchain 1.0 implementations, to a comparison between implementations and mechanisms of Blockchain 1.0, Blockchain 2.0 and later versions, so as to rationalize why the decentralized solutions should be developed with existing Blockchain 2.0 implementations, and why blockchain technology should be utilized to decentralize current solutions implemented in supply chain industry with related work of blockchain-enabled solutions also elaborated in this research.

The overview of NAS, one of the anti-counterfeiting and traceability solutions currently implemented in supply chain industry, is explained with a set of security analysis also performed against it, ranging from asset analysis to threat analysis, where NFC tags, transaction history data and the backend database handling data storage and processing with different microservices involved in different operations required, are amongst the most probable system components susceptible to security risks based on the \emph{CIA methodology} applied. Regarding the threat analysis, there are mainly two categories of threats, namely \emph{physical NFC tag threats} and \emph{system threats}. These indicate the fact that tag-cloning, tag-disabling, tag's data modification, man-in-the-middle relay attack, spoofing attack on product data, and denial-of-service are amongst the most probable and risky threats to NAS as well as other existing anti-counterfeiting and traceability systems built on top of any software system component and wireless tag communication technology.

The discussion of research results indicated some potential opportunities, which are exactly the potential advantages of decentralizing the existing supply chain anti-counterfeiting solutions where it has been maintained merely by product producers or in the case of NAS - the winemakers. The opportunities, such as (1) the improved data integrity with a decentralized product record management with state transitions on product records only accepted if its respective transaction is validated and the block is mined on the blockchain network, (2) the strengthen security considerations on data of product records processed with multilayered validations in place, and (3) the high availability on system components and data of product records enabled by the decentralized system components, including the blockchain network to prevent from possible system downtime and unavailability.

There are also potential concerns on actual development of the decentralized solutions, namely (1) the manageable system integration model on whether it could promote adoptions amongst the industry participants, (2) degree of decentralization which could be contradicting the performance and throughput of processing product records, (3) the limited scalability attributed by the extra decentralized processes and multilayered validations on processing product data, (4) the security vulnerabilities attributed by lack of proper key management and rotation mechanism, and (5) the privacy concern on true identities of targeted entities being exposed, which could be mapped basing on a public address of repetitive uses, by other malicious entities running instances of the decentralized solutions, and even nodes on the same blockchain network. A fundamental set of system requirements for the \emph{Decentralized NFC-Enabled Anti-Counterfeiting System (dNAS)}, which is proposed to decentralize NAS of centralized architecture, is therefore defined, according to the potential concerns discussed, to set requirements for the actual development and implementation of dNAS to take place in a separate research where the proposed system use cases, architecture and implementation will be elaborated with explanation on why the proposed decentralized solution could combat the existing challenge of product counterfeiting in supply chain industry. 

\newpage
\bibliographystyle{ieeetr}

\begin{IEEEbiography}[{\includegraphics[width=1.1in,height=1.25in,clip,keepaspectratio]{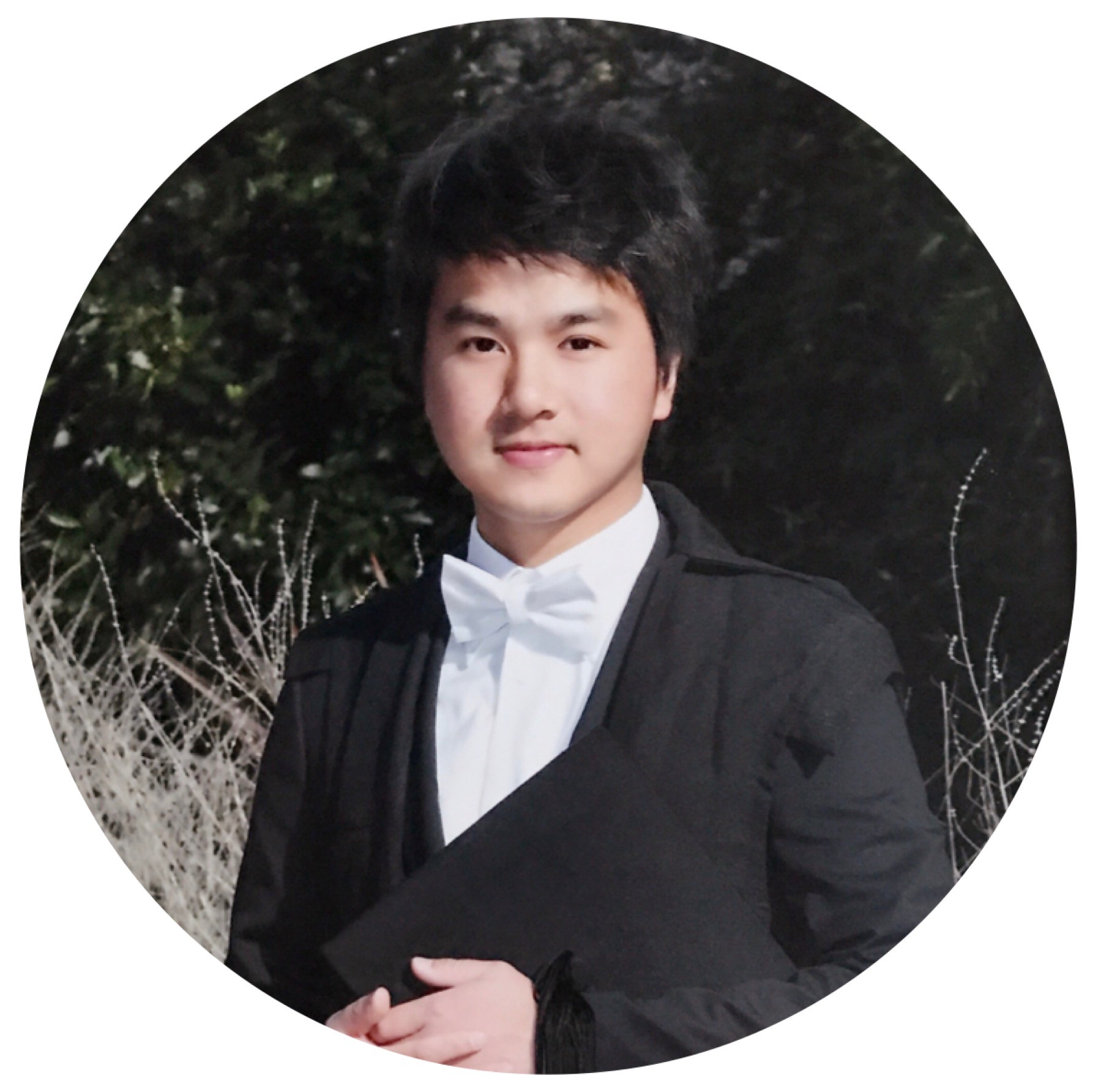}}]{Mr. Neo C.K. Yiu IEEE}
is a computer scientist and software architect specialized in developing decentralized and distributed software solutions for industries. Neo is currently the Lead Software Architect of Blockchain and Cryptography Development at De Beers Group on their end-to-end traceability projects across different value chains with the Tracr™ initiative. Formerly acting as the Director of Technology Development at Oxford Blockchain Society, Neo is currently a board member of the global blockchain advisory board at EC-Council. Neo received his MSc in Computer Science from University of Oxford and BEng in Logistics Engineering and Global Supply Chain Management from The University of Hong Kong.
\end{IEEEbiography}
\vfill

\newpage
\appendices
\section{Security Analysis on NAS}
\subsection{Asset Analysis of NAS} \label{a1}
\begin{figure*}[h]
    \centering
    \captionsetup{justification=centering}
    \includegraphics[width=0.98\textwidth]{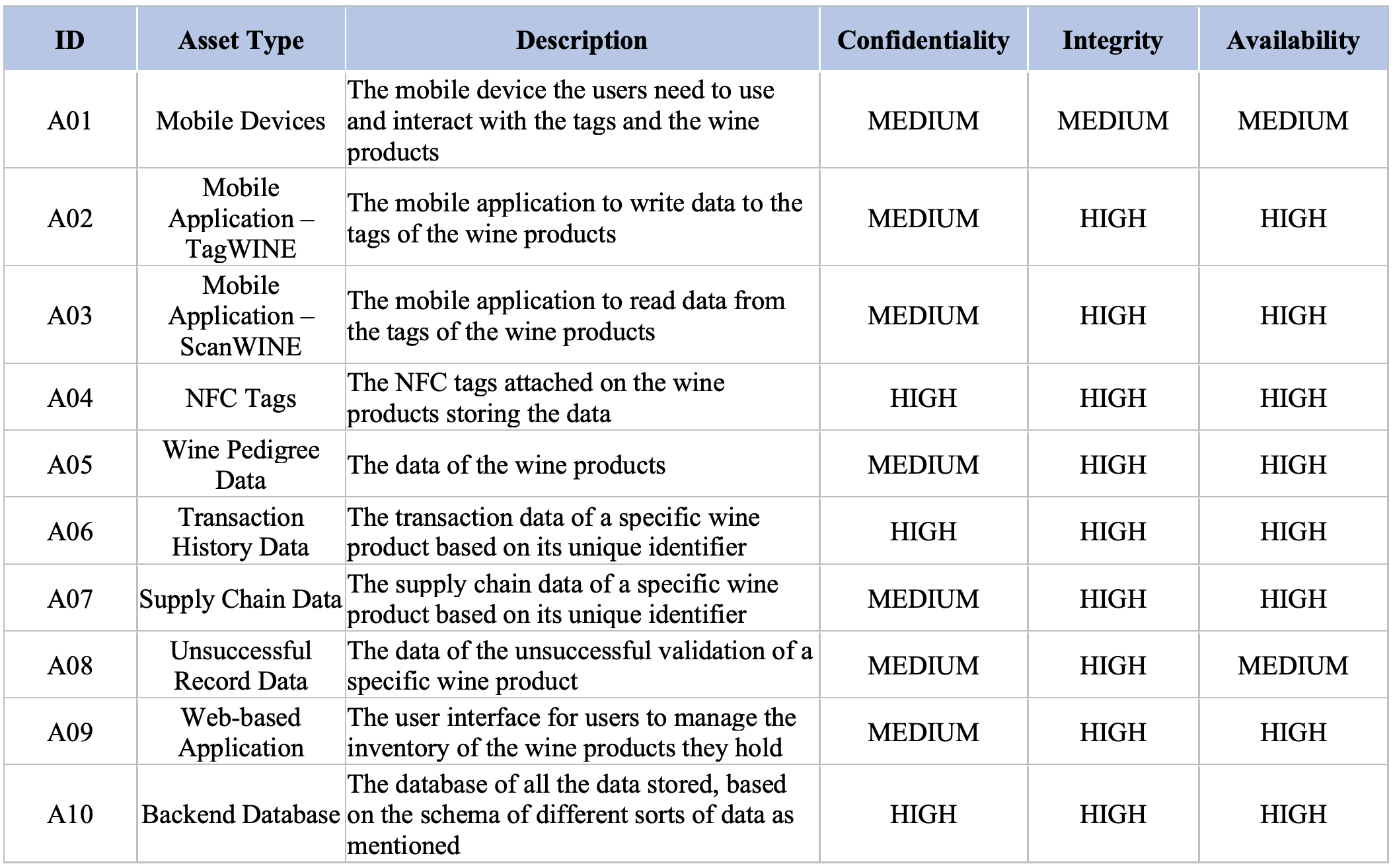}
    \caption{\textit{Asset Analysis of NAS}}
    \label{fig:assetanalysis}
\end{figure*}

The Asset Analysis of NFC-Enabled Anti-Counterfeiting System (NAS) is demonstrated in \textit{Fig.~\ref{fig:assetanalysis}}.

\subsection{Threat Analysis of NAS} \label{a2}
\begin{figure*}[h]
    \centering
    \captionsetup{justification=centering}
    \includegraphics[width=1.07 \textwidth]{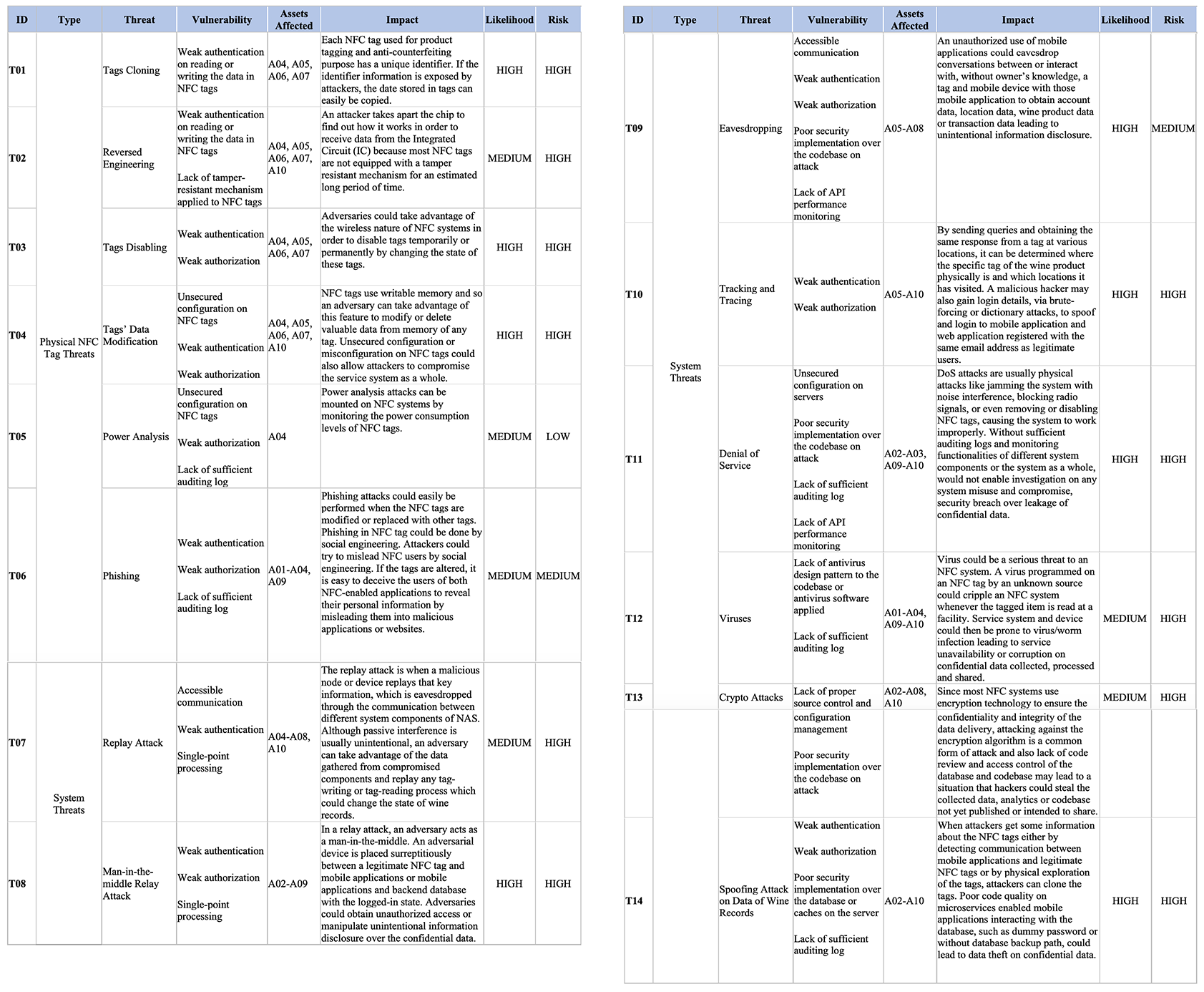}
    \caption{\textit{Threat Analysis of NAS}}
    \label{fig:threatanalysis}
\end{figure*}

The Threat Analysis of NFC-Enabled Anti-Counterfeiting System (NAS) is demonstrated in \textit{Fig.~\ref{fig:threatanalysis}}.

\section{Proposed System Requirement of dNAS}
\subsection{Fundamental System Requirements of dNAS} \label{b1}
\begin{figure*}[h]
    \centering
    \captionsetup{justification=centering}
    \includegraphics[height=1.3 \textwidth]{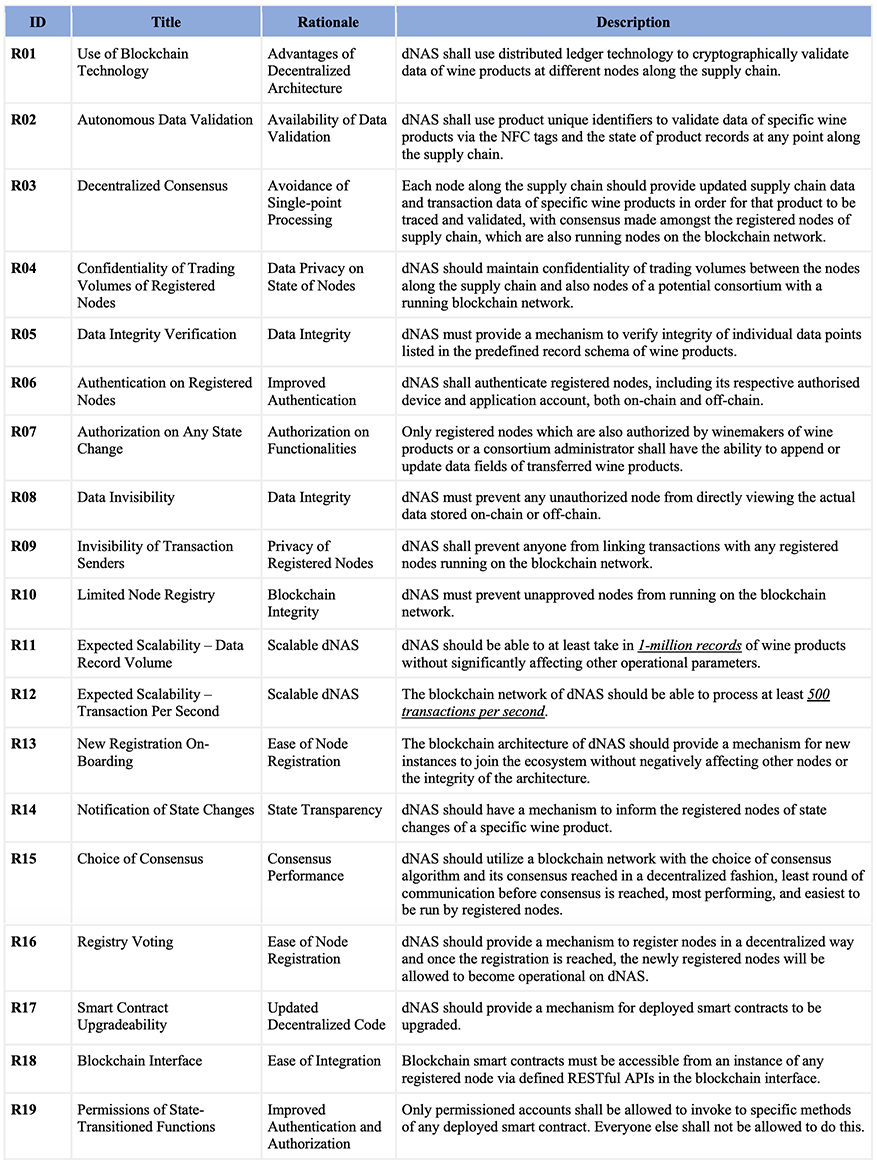}
    \caption{\textit{System Requirements of Proposed dNAS}}
    \label{fig:systemreq}
\end{figure*}

The System Requirements of the proposed Decentralized NFC-Enabled Anti-Counterfeiting System (dNAS) are demonstrated in \textit{Fig.~\ref{fig:systemreq}}.

\end{document}